\documentclass[journal]{IEEEtran}
\usepackage{cite}

\usepackage{subfigure}
\ifCLASSINFOpdf
\usepackage[pdftex]{graphicx}
  \graphicspath{{../pdf/}{../jpeg/}}
  \DeclareGraphicsExtensions{.pdf,.jpeg,.png}
\else
  \usepackage[dvips]{graphicx}
  \graphicspath{{../eps/}}
  \DeclareGraphicsExtensions{.eps}
\fi
\usepackage{amsmath}
\usepackage{cases}
\usepackage{stfloats} 
\usepackage{amsfonts}
\usepackage{subeqnarray}
\usepackage{longtable}
\usepackage{supertabular}
\usepackage{setspace}
\usepackage{multirow}
\usepackage{float}
\usepackage[ruled,linesnumbered]{algorithm2e}
\makeatletter
\newcommand{\nosemic}{\renewcommand{\@endalgocfline}{\relax}}% Drop semi-colon ;
\newcommand{\dosemic}{\renewcommand{\@endalgocfline}{\algocf@endline}}% Reinstate semi-colon ;
\let\oldnl\nl% Store \nl in \oldnl
\newcommand{\nonl}{\renewcommand{\nl}{\let\nl\oldnl}}% Remove line number for one line
\makeatother
\usepackage[export]{adjustbox} 
\usepackage{booktabs}
\usepackage{setspace}
\usepackage{xcolor}
\usepackage{mathrsfs}
\usepackage{amsmath}
\usepackage{array}
\usepackage{amssymb}
\usepackage{amsthm}
\usepackage{microtype}
\usepackage{url}
\usepackage{amsfonts,amssymb}
\usepackage{dsfont}
\usepackage{mathtools}

\allowdisplaybreaks
\begin{document}
\setstretch{0.89}
\title{\textls[-25]{Chance-constrained OPF: A Distributed Method with Confidentiality Preservation}}
\author{Mengshuo~Jia,~\IEEEmembership{Member,~IEEE,} 
Gabriela~Hug,~\IEEEmembership{Senior~Member,~IEEE}, 
Yifan~Su,~\IEEEmembership{Student~Member,~IEEE}, and
Chen~Shen,~\IEEEmembership{Senior~Member,~IEEE}

%\thanks{M. Jia and C. Shen are with the State Key Laboratory of Power Systems, Tsinghua University, 100084 Beijing, China. Y. Wang and G. Hug are with the Power Systems Laboratory, ETH Zurich, 8092 Zurich, Switzerland.}
}
        
%\thanks{This work was supported in part by the Joint Funds of the National Natural Science Foundation of China under Grant U1766206 (Correspondence to Chen Shen).}
%\thanks{M. Jia, C. Shen and Z. Wang are affiliated with the State Key Laboratory of Power Systems, Department of Electrical Engineering, Tsinghua University, Beijing 100084, China (e-mail addresses: jms16@mails.tsinghua.edu.cn, shenchen@mail.tsinghua.edu.cn,
%    wangzhaojian@mail.tsinghua.edu.cn).}% <-this % stops a space
% \thanks{Manuscript received April 19, 2005; revised August 26, 2015.}

%\markboth{Submitted to IEEE Trans. Smart Grid}%
\maketitle

\begin{abstract}
	Given the increased percentage of wind power in power systems, chance-constrained optimal power flow (CC-OPF) calculation, as a means to take wind power uncertainty into account with a guaranteed security level, is being promoted. Compared to the local CC-OPF within a regional grid, the global CC-OPF of a multi-regional interconnected grid is able to coordinate across different regions and therefore improve the economic efficiency when integrating high percentage of wind power generation. In this global problem, however, multiple regional independent system operators (ISOs) participate in the decision-making process, raising the need for distributed but coordinated approaches. Most notably, due to regulation restrictions, commercial interest, and data security, regional ISOs may refuse to share confidential information with others, including generation cost, load data, system topologies, and line parameters. But this information is needed to build and solve the global CC-OPF spanning multiple areas. To tackle these issues, this paper proposes a distributed CC-OPF method with confidentiality preservation, which enables regional ISOs to determine the optimal dispatchable generations within their regions without disclosing confidential data. This method does not require parameter tunings and will not suffer from convergence challenges. Results from IEEE test cases show that this method is highly accurate. 
	
	% Meanwhile, the price brought by the distributed and confidentiality-preserving designs in this method is totally affordable.
\end{abstract}
%Although the historical data of renewable generations could be assumed as publicly known
\begin{IEEEkeywords}
  Wind power, chance constraint, OPF, distributed computing, confidentiality preservation
\end{IEEEkeywords}
\IEEEpeerreviewmaketitle

\section{Introduction}

\IEEEPARstart{G}{iven} the carbon neutrality targets of more than 100 countries \cite{van_soest_net_zero_2021}, wind power, as a major renewable source, has seen rapid development in the past decade and will likely continue to grow in the near future. However, such rapid development may cause significant wind power curtailment if not well integrated. Interconnection among regional power grids provides an effective means to reduce curtailment as connected regional grids can support each other. Most importantly, if the global optimal power flow (OPF) calculations are carried out for the whole interconnected grid, the overall energy security and economic efficiency can be improved \cite{8031400}. However, the uncertainty introduced by renewable generation requires formulating a stochastic OPF problem. Chance constraints, which can limit the risk brought by wind power uncertainty, offer an intuitive way to formulate and solve the stochastic OPF problem \cite{8017474}. All the above motivates the investigation of the chance-constrained OPF (CC-OPF) of multi-regional interconnected grids (MIGs).

% is worthy of investigation, due to its ability to decrease wind power curtailment given wind power fluctuation. 

In previous works, a wide variety of methods have been developed to solve CC-OPF problems. Each of these methods consists of two critical steps, namely (i) the approximation step and (ii) the conversion step. The approximation step transforms the AC power flow model into a linear one \cite{lubin_robust_2016,roald_corrective_2017,zhang_distributionally_2017,schmidli_stochastic_2016,lubin_chance_2019,baker_efficient_2017,baker_joint_2019,9400744}. This step seems to be unavoidable, as a linear model can not only convert the OPF into a convex problem, but also yield an explicit mapping relationship between power flows and power injections, which is the foundation for the following conversion step. Among all the approximation techniques used, the DC power flow is the most commonly-used approach \cite{lubin_robust_2016,roald_corrective_2017,zhang_distributionally_2017,9400744}. Other techniques, say, linearization around the operating point \cite{schmidli_stochastic_2016,lubin_chance_2019} and full linearization of the whole AC model \cite{baker_efficient_2017,baker_joint_2019}, are also adopted for better accuracy. 

The conversion step converts chance constraints into their deterministic versions. According to the chance constraint type, the conversion step can be further divided into two categories, namely converting single chance constraints (SCCs) \cite{lubin_robust_2016,roald_corrective_2017,zhang_distributionally_2017,schmidli_stochastic_2016,lubin_chance_2019}, and converting joint chance constraints (JCCs) \cite{baker_efficient_2017,baker_joint_2019,9400744}. The SCC restricts the violation probability of a single system state (e.g., a line flow) whereas a JCC restricts the combined violation probability of multiple system states simultaneously. While JCCs provide a stronger guarantee of the overall system security compared to SCCs \cite{9122389}, a number of existing methods \cite{baker_efficient_2017,baker_joint_2019,9400744} are capable to decompose JCCs into SCCs. Therefore, converting SCCs into deterministic constraints is the more fundamental step. 

Wind power uncertainty is usually assumed to follow the normal distribution \cite{lubin_robust_2016,roald_corrective_2017,zhang_distributionally_2017,schmidli_stochastic_2016,lubin_chance_2019}. Consequently, each SCC can be equivalently converted into its deterministic version through quantile calculation (i.e., the inverse calculation of the normal cumulative distribution function) \cite{lubin_robust_2016,roald_corrective_2017,zhang_distributionally_2017,schmidli_stochastic_2016,lubin_chance_2019}. However, the hypothesis of following a normal distribution is not suitable for non-Gaussian wind power uncertainty \cite{wang_chance_constrained_2017}. To address this problem, a Gaussian mixture model (GMM), which is flexible enough to capture different distribution characteristics (including bias, heavy tail, and multi-peak) \cite{ge_parameter_2018}, is adopted by many existing works to formulate the probability distribution of wind power uncertainty \cite{wang_chance_constrained_2017,9400744,wang_adjustable_2018}. Conveniently, the quantile-based conversion strategy is still applicable to GMMs and therefore, any distribution model can be considered.
% \vspace{-5pt}
\begin{table*}[ht]
  \centering
  \footnotesize
  \caption{Confidential Data of Each Regional Grid and the Reason Why the Data Cannot Be Leaked}
  \renewcommand{\arraystretch}{1}
  \linespread{0.5}
  \begin{tabular}{p{4.5cm} p{12.5cm}}
    \toprule
    \bfseries Confidential Data
          & \bfseries Reason Why the Data Cannot Be Leaked                        \\
    \midrule 
    \multicolumn{1}{m{4.5cm}}{Generation cost of dispatchable generators}   & \multicolumn{1}{m{12.5cm}}{First, once other stakeholders, especially the adversaries, know the generation cost of the dispatchable generators within a regional grid, they are able to cause economical damage \cite{9097933}, or gain more benefits in the market \cite{9152445}; Second, in some mature electricity markets, e.g., Pennsylvania—New Jersey—Maryland (PJM), ISOs are prohibited from disclosing commercially sensitive information to others in general cases \cite{8031400,amended2017restated}.}  \\
	\midrule 
    \multicolumn{1}{m{4.5cm}}{Current outputs and available capacities of dispatchable generators}   & \multicolumn{1}{m{12.5cm}}{In PJM, for example, the current outputs and available capacities of generators are regarded as confidential information. If ISOs from other regions want to access these data, they must apply to PJM first \cite{generator_confi}. The application will only be approved if this request is necessary. }\\
	\midrule 
    \multicolumn{1}{m{4.5cm}}{Load data}  &  \multicolumn{1}{m{12.5cm}}{Once the load data of an area is shared with others, the number of eavesdroppers that may access this load data will increase. This simultaneously enhances the risk of the load-information-based attack to this area, e.g., the attack on the automatic generation control system \cite{6740883}, as knowing the load data is one of the main prerequisites to implementing such an attack.}  \\
	\midrule 
    \multicolumn{1}{m{4.5cm}}{Grid topology and line parameters} &  \multicolumn{1}{m{12.5cm}}{Sharing the grid topology and line parameters of a region might increase the risk of an  attack on the state estimation of this region \cite{liu2011false}. Such an attack will greatly influence the decision of opening or closing lines, changing the position of transformer taps, etc. \cite{chatterjee2017review}. } \\
    \bottomrule
  \end{tabular}
  \label{tab:data barrier reason}
\end{table*}

Although the research on CC-OPF is already diverse, all the aforementioned CC-OPF methods have to be implemented in a centralized manner. In a MIG, however, multiple regional independent system operators (ISOs) participate in the decision-making, rendering centralized algorithms unsuitable. First, it is unlikely that there is a central operator with access to all the data of the regional grids in a MIG  \cite{6689354,8031400}. Second, even in cases where a central authority (or say an upper-level operator) does exist, e.g., the National Power Dispatch Center of China (NPDCC) that coordinates regional ISOs within the country, the OPF of an interconnected grid is still unlikely to be implemented centrally. This is because different control centers have a clear division of  functionality/responsibility \cite{7041236}. Specifically, in the MIG of China, the NPDCC is in charge of  scheduling the transmission plan of tie lines between regions according to the power transaction contracts, while each regional ISO is responsible for deciding the generation dispatch within its own area using the transmission plan as boundary condition. No one is authorized to make decisions for others, i.e., the NPDCC and regional ISOs function independently \cite{7041236}. Such a hierarchical management structure makes the global OPF of the MIG unsuitable and thus traditional centralized OPF methods not applicable.

% Clearly, in this hierarchical management structure, the transmission plan and the generation dispatch within each area are determined sequentially instead of simultaneously, which goes against the basic idea of 

% making neither the transmission plan nor the regional generation dispatch global optimal. In another word, the global OPF of the MIG in China is still not implemented centrally even if a central authority exists, causing centralized CC-OPF algorithms not applicable.

Since centralized methods are not suitable for the CC-OPF of MIGs, distributed methods should be considered. Despite the fact that there is a vast literature on distributed OPF approaches, to the best of the authors' knowledge, there is no distributed method for CC-OPF problems. The reason might lie in the unique nature of chance constraints. E.g., the coefficients in a line flow SCC are mainly dependent on the topological connections between that particular line and the dispatchable generators all over the whole grid. That is to say, these coefficients are not locally obtainable unless the global information of the whole grid is known, which again is not the case in a MIG. Consequently, developing a distributed method for CC-OPF is more complicated than merely proposing a distributed optimization algorithm. So far, how to solve the CC-OPF problem in a distributed manner remains an open question. Accordingly, this paper aims to develop a distributed CC-OPF method for MIGs.

% Although the distributed method is more applicable compared to the centralized approach, 
It should be noted that information exchange between data owners is always necessary in distributed methods. Most importantly, such information exchange may disclose the confidential data of data owners (e.g., regional ISOs in this paper) \cite{9097933}. However, the leakage of confidential information is unacceptable for regional ISOs \cite{9097933,9152445,8031400,amended2017restated,generator_confi,6740883,liu2011false,chatterjee2017review}. For details, please refer to Table \ref{tab:data barrier reason}, which offers the types of confidential data for a regional grid as well as the reason why such data should not be disclosed. Hence, regional ISOs actually need a confidentiality-preserving distributed CC-OPF method (instead of a pure distributed approach), which not only can enable regional ISOs to solve their global CC-OPF in a distributed way, but also can protect the confidential data of each ISO. 

% which not only can enable regional ISOs to solve their global CC-OPF in a distributed way, but also can protect the confidential data of each area from disclosure, is needed for practical use. 

% \vspace{-1pt}
To this end, this paper, on the basis of the transformation-based encryption (TE) technique \cite{li2013privacyv,li2013privacyh,8031400}, proposes a confidentiality-preserving distributed CC-OPF method. The advantage of the TE techniques is that it not only can solve optimization problems in a confidentiality-preserving manner, but also does not require parameter tunings or iterative calculations \cite{li2013privacyv,li2013privacyh,8031400}. In contrast, other distributed optimization algorithms, such as the alternating direction method of multipliers (ADMM), normally need sophisticated parameter tuning approaches to achieve satisfactory convergence. Yet, parameter tuning is usually problem-specific without a universal rule, and improper tuning may lead to divergence \cite{8932561,8408802}. It is worth mentioning that, although the TE technique has appealing features, there are still some unsolved challenges if using this technique for CC-OPF. Specifically, the TE technique is only proven to be effective for linear programming so far, but the CC-OPF problem is nonlinear. In addition, the TE technique still faces the issue that the coupled coefficients in SCCs are locally unobtainable for regional ISOs. To address these challenges, this paper first proves two theorems to ensure that the TE technique can be adopted for the nonlinear CC-OPF problem. Then, this paper reveals that computing the coupled coefficients in SCCs corresponds to solving a system of linear equations. Accordingly, a fast distributed algorithm for solving linear equation systems is proposed, which enables regional ISOs to obtain the coupled coefficients in SCCs without revealing their confidential data. Finally, this paper proposes a distributed CC-OPF method with confidentiality preservation. Therefore, the contributions of this paper are as follows:
\begin{enumerate}
%	\item An extreme value judgment method is leveraged, which guarantees that binding constraints will never be classified as non-binding.
	\item Prove the TE technique's adaptability to quadratic programming, ensuring that this technique can find the centralized optimal solution of the CC-OPF problem.
	\item Propose a fast confidentiality-preserving distributed algorithm for solving linear-equations-system, which has both higher accuracy and higher efficiency than the corresponding state-of-the-art methods.
	\item Propose a distributed CC-OPF method with confidentiality preservation. 
\end{enumerate}

In the following, Section II first formulates the CC-OPF problem for a MIG, and then discusses the privacy issues when building and solving this problem. After that, Section III briefly revisits the TE technique, and also explains the unsolved challenges if applied to the CC-OPF. These challenges are further addressed in Section IV. Consequently, Section V presents the distributed CC-OPF algorithm with confidentiality preservation. Finally, case studies are performed in Section VI and Section VII concludes this paper.

\section{CC-OPF Model for MIG}
In this section, a CC-OPF problem is formulated first. Then, the SCCs in this problem are converted into their deterministic versions, accordingly generating a compact formulation of this model. Finally, the confidentiality issues when formulating and solving this problem are revealed.

\subsection{CC-OPF}
We start by providing a description of the notations used in the CC-OPF formulation. Let $\mathcal{M}$ be the set of regions in the grid. The dispatchable generator set, non-dispatchable wind farm set, load set, and transmission line set of the whole grid are represented by $\mathcal{G}$, $\mathcal{W}$,  $\mathcal{D}$, and $\mathcal{L}$, respectively. Correspondingly, the sets of generators, wind farms, loads, and lines in area $n$ are denoted by $\mathcal{G}_n$, $\mathcal{W}_n$, $\mathcal{D}_n$, and $\mathcal{L}_n$, respectively. This paper uses $\mathcal{T}$ as the set of time slots, and uses $|\cdot|$ to express the cardinality calculation of any set. Besides, the active power output of dispatchable generators, active wind power outputs, reactive wind power outputs, active loads, reactive loads, and active line flows in area $n$ at time $t$ are respectively denoted by $\boldsymbol{g}_{n,t} \in \mathfrak{R}^{|\mathcal{G}_n|} $, $\boldsymbol{w}_{n,t} \in \mathfrak{R}^{|\mathcal{W}_n|} $, $\boldsymbol{\widehat{w}}_{n,t} \in \mathfrak{R}^{|\mathcal{W}_n|} $, $\boldsymbol{d}_{n,t} \in \mathfrak{R}^{|\mathcal{D}_n|} $, $\boldsymbol{\widehat{d}}_{n,t} \in \mathfrak{R}^{|\mathcal{D}_n|} $, and $\boldsymbol{l}_{n,t} \in \mathfrak{R}^{|\mathcal{L}_n|} $. We further introduce $N_a=|\mathcal{M}|$, $T=|\mathcal{T}|$, $H_n=|\mathcal{G}_n|\times |\mathcal{T}|$, and $H=|\mathcal{G}|\times |\mathcal{T}|$.

The objective of the CC-OPF is to minimize the overall generation cost across the entire grid given the non-dispatchable wind power: 
\begin{align}
	\min_{\boldsymbol{g}_{n,t} } \ \mathcal{F}_0 =  \sum_{t \in \mathcal{T} }\sum_{n \in \mathcal{M} } \frac{1}{2}\,\boldsymbol{g}_{n,t}^\top\, \boldsymbol{\mathcal{I}}_n\,\boldsymbol{g}_{n,t} \  + \sum_{t \in \mathcal{T} }\sum_{n \in \mathcal{M} }\boldsymbol{\mathcal{K}}_n^\top\boldsymbol{g}_{n,t}  \label{eq:Original_obj}
\end{align}
where $\boldsymbol{\mathcal{I}}_n \in \mathfrak{R}^{|\mathcal{G}_n|\times |\mathcal{G}_n|}$ is a diagonal matrix composed of the quadratic cost coefficients of dispatchable generators in region $n$, whose non-zero elements are all positive, rendering \eqref{eq:Original_obj} strictly convex. Vector $\boldsymbol{\mathcal{K}}_n \in \mathfrak{R}^{|\mathcal{G}_n|}$ consists of the linear cost coefficients of dispatchable generators in region $n$, whose elements are assumed to be positive.

Due to the non-dispatchable wind power, the supply-demand constraint at time $t$ is formulated as an SCC, as in \cite{8254387,9146814}:  
\begin{align}
	\mathbb{P}\Big\{\sum\nolimits_{n \in \mathcal{M}} \mathds{1}^\top_{|\mathcal{G}_n|} &\boldsymbol{g}_{n,t} \  + \sum\nolimits_{n \in \mathcal{M}} \mathds{1}^\top_{|\mathcal{W}_n|} \boldsymbol{w}_{n,t} \notag \\
  & - \sum\nolimits_{n \in \mathcal{M}} \mathds{1}^\top_{|\mathcal{D}_n|}\boldsymbol{d}_{n,t} \geq 0 \Big\} \geq 1-\epsilon_{b} \label{eq:supply-demand_dis}
\end{align}
This constraint enforces the supply to be greater than or equal to the demand with a probability of at least $1-\epsilon_{b}$. Vector $\mathds{1}_{|\mathcal{G}_n|}$ is an all-ones vector of size $|\mathcal{G}_n|$. Similarly, $\mathds{1}_{|\mathcal{W}_n|}$ and $\mathds{1}_{|\mathcal{D}_n|}$ are also all-ones vectors with appropriate dimensions. Besides, the wind power outputs $\boldsymbol{w}_{n,t}$ are random variables whereas $\boldsymbol{d}_{n,t}$ is constant, that is, this paper ignores the uncertainty of load because it is relatively small compared to wind power uncertainty \cite{7565532}. 

Constraints for the dispatchable generators in area $n$ at time $t$ include the capacity and ramp rate constraints:
\begin{gather}
  \boldsymbol{g}_n^- \leq  \boldsymbol{g}_{n,t} \leq \boldsymbol{g}_n^+, \ \ \boldsymbol{r}_n^- \leq \boldsymbol{g}_{n,t+1} - \boldsymbol{g}_{n,t}   \leq \boldsymbol{r}_n^+ \notag
\end{gather}
where $\boldsymbol{g}_n^+$ and $\boldsymbol{g}_n^-$ represent the upper and lower generation bounds of generators in area $n$. Similarly, $\boldsymbol{r}_n^+$ and $\boldsymbol{r}_n^-$ are the ramp rate bounds of the generators in area $n$. All of these bounds are vectors in $\Re^{|\mathcal{G}_n|\times 1}$.

Constraints for system states (e.g., nodal voltages or line flows) are also formulated as SCCs. Let $s_{n,i,t}$ denote state $i$ in area $n$ at time $t$, which could be either a nodal voltage or a line flow. The constraint for $s_{n,i,t}$ is given as follows:  
\begin{align}
	\mathbb{P}\left\{  s_{n,i,t} \leq s_{n,i}^+\right\} \geq 1-\alpha_S, \ \ \forall i \in \mathcal{S}_n, \ \forall n \in \mathcal{M} \label{eq:state}
\end{align}
where $s_{n,i}^+ \in \mathfrak{R}$ is the upper bound of $s_{n,i,t}$, and set $\mathcal{S}_n$ consists of the states in region $n$. Constraint \eqref{eq:state} enforces $s_{n,i,t}$ to be within its bound with a probability of at least $1 - \alpha_S$. 

% For the sake of brevity, this constraint does not take the lower limit of $s_{n,i,t}$ into account. Yet even if the lower limit is considered, the essence of the CC-OPF model remains the same --- merely increasing the number of SCCs of the same type.

% , making the effectiveness of the methods proposed in the following sections unchanged.

% it merely increases the number of constraints, which does not affect the essence of the CC-OPF model, nor affect 

% To solve the above CC-OPF, one should first convert the SCCs into their deterministic ones, as describing in the next subsection. 

\subsection{Conversion of SCCs}
Converting the SCCs in \eqref{eq:supply-demand_dis} and \eqref{eq:state} into their deterministic versions is the prerequisite for solving the CC-OPF problem. To this end, this paper first adopts a linear but sufficiently accurate power flow model, namely the decoupled linearized power flow (DLPF) model proposed in \cite{7782382}, whose performance has been verified in \cite{8269410,8413105}. Accordingly, $s_{n,i,t}$, i.e. line flow or voltage, can be expressed as a linear function of power injections across regions:
 \begin{align}
	\label{eq:state_DLPF}
	s_{n,i,t} = & \sum\nolimits_{m \in \mathcal{M}}  \boldsymbol{\mathcal{O}}_{n,i,m}^\top\boldsymbol{w}_{m,t} + \sum\nolimits_{m \in \mathcal{M}} \boldsymbol{\widehat{\mathcal{O}} }_{n,i,m}^\top\boldsymbol{\widehat{w}}_{m,t} \ + \notag \\
   & \sum\nolimits_{m \in \mathcal{M}} \boldsymbol{\mathcal{Y}}_{n,i,m}^\top\boldsymbol{d}_{m,t} \, +\, \sum\nolimits_{m \in \mathcal{M}} \boldsymbol{\widehat{\mathcal{Y}}}_{n,i,m}^\top\boldsymbol{\widehat{d}}_{m,t} \ + \notag \\
   &  \sum\nolimits_{m \in \mathcal{M}} \boldsymbol{\mathcal{U}}_{n,i,m}^\top \boldsymbol{g}_{m,t} \ + \xi_{n,i,t}
\end{align}
% \begin{align}
% 	\label{eq:state_DLPF}
% 	s_{n,i,t} = &  \sum_{m \in \mathcal{M}} \boldsymbol{\mathcal{U}}_{n,i,m}^\top \boldsymbol{g}_{m,t} \ + \sum_{m \in \mathcal{M}} \boldsymbol{\mathcal{U}}_{n,i,m}^\top \boldsymbol{g}_{m,t}^0 \ + \notag \\
%   & \sum_{m \in \mathcal{M}}  \boldsymbol{\mathcal{O}}_{n,i,m}^\top\boldsymbol{w}_{m,t} + \sum_{m \in \mathcal{M}} \boldsymbol{\widehat{\mathcal{O}} }_{n,i,m}^\top\boldsymbol{\widehat{w}}_{m,t} \ + \notag \\
% 	 & \sum_{m \in \mathcal{M}}  \boldsymbol{\mathcal{O}}_{n,i,m}^\top\boldsymbol{w}_{m,t}^0 + \sum_{m \in \mathcal{M}} \boldsymbol{\widehat{\mathcal{O}} }_{n,i,m}^\top\boldsymbol{\widehat{w}}_{m,t}^0 \ + \notag \\
%    & \sum_{m \in \mathcal{M}} \boldsymbol{\mathcal{Y}}_{n,i,m}^\top\boldsymbol{d}_{m,t} \, +\, \sum_{m \in \mathcal{M}} \boldsymbol{\widehat{\mathcal{Y}}}_{n,i,m}^\top\boldsymbol{\widehat{d}}_{m,t} 
% \end{align}
where $\boldsymbol{\mathcal{U}}_{n,i,m} \in \mathfrak{R} ^{|\mathcal{G}_m| }$ consists of the mapping coefficients (derived from the DLPF model) that map $\boldsymbol{g}_{m,t}$ to $s_{n,i,t}$. Similarly, $\boldsymbol{\mathcal{O}}_{n,i,m} \in \mathfrak{R} ^{|\mathcal{W}_m|} $, $\boldsymbol{\widehat{\mathcal{O}}}_{n,i,m} \in \mathfrak{R} ^{|\mathcal{W}_m|} $, $\boldsymbol{\mathcal{Y}}_{n,i,m} \in \mathfrak{R} ^{|\mathcal{D}_m|} $, and $\boldsymbol{\widehat{\mathcal{Y}}}_{n,i,m} \in \mathfrak{R} ^{|\mathcal{D}_m|} $ also consist of the corresponding mapping coefficients. Note that $\xi_{n,i,t}$ is a constant value computed from all known system states, e.g., the voltage magnitude/angle of the slack node, voltage magnitudes of $PV$ nodes, etc. For more details regarding the DLPF model, please see \cite{7782382}.

Using \eqref{eq:state_DLPF}, the SCCs in \eqref{eq:supply-demand_dis} and \eqref{eq:state} can be converted into deterministic constraints using the quantile calculation method proposed in \cite{wang_chance_constrained_2017}. For \eqref{eq:supply-demand_dis}, the resulting deterministic constraint is given by
\begin{align}
-\sum\nolimits_{n \in \mathcal{M}} \mathds{1}^\top_{|\mathcal{G}_n|}\boldsymbol{g}_{n,t}\leq \upsilon_t \label{eq:supply-demand-deter}
\end{align}
where
\begin{align}
	\upsilon_t & = \mathbb{P}_{\omega_t}^{-1}(\epsilon_{b}) - \sum\nolimits_{n \in \mathcal{M}} \mathds{1}^\top _{|\mathcal{D}_n|} \boldsymbol{d}_{n,t}  \notag \\
	\omega_t & =\sum\nolimits_{n \in \mathcal{M}} \mathds{1}^\top_{|\mathcal{W}_n|} \boldsymbol{w}_{n,t} \notag
\end{align}
and $\mathbb{P}_{\omega_t}^{-1}(\epsilon_{b})$ is the $\epsilon_{b}$-quantile of $\omega_t$.

Similarly, the SCC in \eqref{eq:state} is transformed into: 
\begin{align}
	\sum\nolimits_{m \in \mathcal{M}} \boldsymbol{\mathcal{U}}_{n,i,m}^\top \boldsymbol{g}_{m,t} \leq \mathcal{J}_{n,i,t}  \label{eq:line constraint}
\end{align}
where
\begin{align}
	\mathcal{J}_{n,i,t}  = & - \sum\nolimits_{m \in \mathcal{M}} \boldsymbol{\mathcal{Y}}_{n,i,m}^\top\boldsymbol{d}_{m,t} - \sum\nolimits_{m \in \mathcal{M}} \boldsymbol{\widehat{\mathcal{Y}}}_{n,i,m}^\top\boldsymbol{\widehat{d}}_{m,t} \notag \\
	& - \xi_{n,i,t} + s_{n,i}^+ -\mathbb{P}_{\Omega_{n,i,t}}^{-1}\left(1-\alpha_S \right) \notag \\
	\Omega_{n,i,t} = & \sum\nolimits_{m \in \mathcal{M}}  \boldsymbol{\mathcal{O}}_{n,i,m}^\top\boldsymbol{w}_{m,t} + \sum\nolimits_{m \in \mathcal{M}} \boldsymbol{\widehat{\mathcal{O}} }_{n,i,m}^\top\boldsymbol{\widehat{w}}_{m,t} \notag
\end{align}
and $\mathbb{P}_{\Omega_{n,i,t}}^{-1}\left(1-\alpha_S\right)$ is the $(1-\alpha_S)$-quantile of $\Omega_{n,i,t}$.

\vspace{-0.2cm}
\subsection{Compact Formulation of the Converted CC-OPF}
After the above conversions, the CC-OPF problem becomes a quadratic programming problem with linear constraints. Its compact formulation, defined as ($\rm P0$), is given as follows: 
\begin{alignat}{2}
	(\rm P0)\ \ \min_{\boldsymbol{x}} \quad &  \mathcal{F}_0 =  \frac{1}{2}\boldsymbol{x}^\top \! \boldsymbol{Q} \boldsymbol{x} + \boldsymbol{c}^\top\!\boldsymbol{x} \notag  \\
	\mbox{s.t.}\quad
	&\boldsymbol{\mathbb{A}} \boldsymbol{x}\leq \boldsymbol{\mathbb{B}} \notag 
\end{alignat}
Matrix $\boldsymbol{Q}$ in the objective function is composed of the quadratic cost coefficients of dispatchable generators:
\begin{align}
	\boldsymbol{Q} & = 
	\left[
	\begin{array}{ccc}
	\boldsymbol{\Lambda}_1  & \cdots & \boldsymbol{0} \\
	\vdots & \ddots & \vdots \\[1mm]
	\boldsymbol{0} &  \cdots & \boldsymbol{\Lambda}_{N_a} \\
	\end{array}
\right] \in \mathfrak{R}^{H\times H }  \notag
\end{align}
where $\boldsymbol{\Lambda}_n$ is detailed by
\begin{equation}
	\boldsymbol{\Lambda}_n = 
	\begin{bmatrix}
		\boldsymbol{\mathcal{I}}_n & \cdots & 0 \\
		\vdots & \ddots & \vdots \\
		0 & \cdots & \boldsymbol{\mathcal{I}}_n \\
	\end{bmatrix} \in \mathfrak{R}^{H_n \times H_n }  \notag
\end{equation}
Furthermore, $\boldsymbol{c}$ in the objective function consists of the linear cost coefficients of dispatchable generators of the whole grid, i.e.
\begin{align}
	\boldsymbol{c} & = 
	\begin{bmatrix}
		\boldsymbol{c}_1^\top & \cdots & \boldsymbol{c}_{N_a}^\top 
	\end{bmatrix}^\top \in \mathfrak{R}^{H} \notag
\end{align}
where $\boldsymbol{c}_n$ is defined as follows:
\begin{align}
	\boldsymbol{c}_n & = 
	\begin{bmatrix}
		\boldsymbol{\mathcal{K}}_{n}^\top & \cdots & \boldsymbol{\mathcal{K}}_{n}^\top
	\end{bmatrix}^\top \in \mathfrak{R}^{H_n } \notag
\end{align}
The decision variable $\boldsymbol{x}$ includes the active dispatchable generations of the whole grid for all time steps:
\begin{align}
	\boldsymbol{x} & =	
	\begin{bmatrix}
		\boldsymbol{x}_1^\top & \cdots & \boldsymbol{x}_{N_a}^\top
	\end{bmatrix}^\top \in \mathfrak{R}^{H} \label{eq:x}
\end{align}
where 
\begin{align}
	\boldsymbol{x}_n & = 
	\begin{bmatrix}
		\boldsymbol{g}_{n,1}^\top & \cdots & \boldsymbol{g}_{n,T}^\top
	\end{bmatrix}^\top \in \mathfrak{R}^{H_n} \label{eq:x_n}
\end{align}

Before describing $\boldsymbol{\mathbb{A}}$ and $\boldsymbol{\mathbb{B}}$ in $(\rm{P0})$, several parameter matrices are first defined, including $\boldsymbol{I}_{n} \in \mathfrak{R}^{|\mathcal{G}_n| \times |\mathcal{G}_n|}$, $\boldsymbol{J}_n \in \mathfrak{R}^{H_n \times H_n}$, $\boldsymbol{E}_n \in \mathfrak{R}^{(H_n -|\mathcal{G}_n|) \times H_n}$, $\boldsymbol{K}_n \in \mathfrak{R}^{T \times H_n}$, $\boldsymbol{\Phi}_n \in \mathfrak{R}^{|\mathcal{S}| \times |\mathcal{G}_n|}$, and $\boldsymbol{U}_n \in \mathfrak{R}^{T|\mathcal{S}| \times H_n}$, where $\mathcal{S} = \bigcup_{n \in \mathcal{M} } \mathcal{S}_n$. Among these parameter matrices, $\boldsymbol{I}_{n}$ is an identity matrix; other parameter matrices are specified as follows:
% \begin{equation}
% 	\boldsymbol{J}_n = 
% 	\left[                 
% 	  \begin{array}{ccc}   
% 		\boldsymbol{I}_n &  \cdots & \boldsymbol{0} \\[1mm]
% 		\vdots &   \ddots & \vdots \\[1mm]
% 		\boldsymbol{0} &  \cdots & \boldsymbol{I}_n \\
% 	  \end{array}
% 	\right] \notag 
% \end{equation}
\begin{align}
	& \boldsymbol{J}_n = 
	\left[                 
	  \begin{array}{ccc}   
		\boldsymbol{I}_n &  \cdots & \boldsymbol{0} \\[1mm]
		\vdots &   \ddots & \vdots \\[1mm]
		\boldsymbol{0} &  \cdots & \boldsymbol{I}_n \\
	  \end{array}
	\right] \notag \\
	& \boldsymbol{E}_{n} = 
		\left[                 
		\begin{array}{ccccc}   
		-\boldsymbol{I}_n & \boldsymbol{I}_n  &\cdots & \boldsymbol{0} & \boldsymbol{0} \\[1mm]
		\vdots & \vdots  & \ddots & \vdots & \vdots \\[1mm]
		\boldsymbol{0} & \boldsymbol{0}  & \cdots &  -\boldsymbol{I}_n & \boldsymbol{I}_n \\
		\end{array}
	\right]  \notag \\
	& \boldsymbol{K}_{n}  = 
		\left[                 
		\begin{array}{ccc}   
		-\mathds{1}^\top_{|\mathcal{G}_n|} & \cdots & \boldsymbol{0}  \\[1mm]
		\vdots & \ddots & \vdots  \\[1mm]
		\boldsymbol{0} &\! \cdots &  -\mathds{1}^\top_{|\mathcal{G}_n|}  \\
		\end{array}
	\right] \notag  \\
	& \boldsymbol{\Phi}_n \!   = \! \left[\boldsymbol{\mathcal{U}}_{1,1,n}\cdots\boldsymbol{\mathcal{U}}_{1,|\mathcal{S}_1|,n} \cdots\boldsymbol{\mathcal{U}}_{{N_a},1,n}\cdots\boldsymbol{\mathcal{U}}_{{N_a},|\mathcal{S}_{N_a}|,n} \right]^\top  \label{eq:phin} \\
	& \boldsymbol{U}_n \!   = \!
	\left[
		\begin{array}{cccc}   
		\boldsymbol{\Phi}_n & \boldsymbol{0} & \cdots & \boldsymbol{0} \\[1mm]
		\boldsymbol{0} & \boldsymbol{\Phi}_n & \cdots & \boldsymbol{0} \\
		\boldsymbol{0} & \boldsymbol{0} &   \ddots & \boldsymbol{0} \\[1mm]
		\boldsymbol{0} & \boldsymbol{0} &   \cdots & \boldsymbol{\Phi}_n \\
		\end{array}
	\right]  \label{eq:Un}
\end{align}

Additionally, the following parameter vectors are introduced:
\begin{align}
	\boldsymbol{G}_n^+ & = \left[\boldsymbol{g}_{n}^{+^\top} \ \ \cdots  \ \ \boldsymbol{g}_{n}^{+^\top} \right]^\top \in \mathfrak{R}^{H_n } \notag \\
	\boldsymbol{G}_n^- & = \left[\boldsymbol{g}_{n}^{-^\top}  \ \ \cdots  \ \ \boldsymbol{g}_{n}^{-^\top} \right]^\top \in \mathfrak{R}^{H_n } \notag \\
	\boldsymbol{R}_n^+ & = \left[\boldsymbol{r}_{n}^{+^\top}  \ \ \cdots  \ \ \boldsymbol{r}_{n}^{+^\top} \right]^\top \in \mathfrak{R}^{(H_n-|\mathcal{G}_n|) } \notag \\
	\boldsymbol{R}_n^- & = \left[\boldsymbol{r}_{n}^{-^\top}  \ \ \cdots  \ \ \boldsymbol{r}_{n}^{-^\top} \right]^\top \in \mathfrak{R}^{(H_n-|\mathcal{G}_n|) }  \notag \\
	\boldsymbol{\Upsilon} & = \left[\upsilon_{1}  \ \ \cdots  \ \ \upsilon_{T} \right]^\top \in \mathfrak{R}^{T} \notag \\
	\boldsymbol{\Delta}_t &	 = \left[\mathcal{J}_{1,1,t}   \cdots   \mathcal{J}_{1,|\mathcal{S}_1|,t}   \cdots   \mathcal{J}_{{N_a},1,t}   \cdots   \mathcal{J}_{{N_a},|\mathcal{S}_{N_a}|,t} \right]^\top \!\! \in \mathfrak{R}^{|\mathcal{S}| } \notag \\
	\boldsymbol{\Delta} &  = \left[\boldsymbol{\Delta}_1^\top   \ \ \cdots  \ \ \boldsymbol{\Delta}_T^\top \right]^\top \in \mathfrak{R}^{T|\mathcal{S}| } \notag
\end{align}

Based on the above parameter matrices and vectors, $\boldsymbol{\mathbb{A}}$ and $\boldsymbol{\mathbb{B}}$ are given by
\begin{align}
	\label{eq:Ab}
	& \boldsymbol{\mathbb{A}} = 
\left[                 
  \begin{array}{ccc}   
	\boldsymbol{J}_1  & \cdots & \boldsymbol{0} \\
	\vdots & \ddots & \vdots \\[1mm]
	\boldsymbol{0} & \cdots & \boldsymbol{J}_{N_a} \\[1mm]
	-\boldsymbol{J}_1  & \cdots & \boldsymbol{0} \\
	\vdots & \ddots & \vdots \\[1mm]
	\boldsymbol{0}  & \cdots & -\boldsymbol{J}_{N_a} \\[1mm]
	\boldsymbol{E}_1 & \cdots & \boldsymbol{0} \\
	\vdots & \ddots & \vdots \\[1mm]
	\boldsymbol{0}  & \cdots & \boldsymbol{E}_{N_a} \\[1mm]
	-\boldsymbol{E}_1 & \cdots & \boldsymbol{0} \\
	\vdots & \ddots & \vdots \\[1mm]
	\boldsymbol{0} & \cdots & -\boldsymbol{E}_{N_a} \\[1mm]
	\boldsymbol{K}_1  &\cdots & \boldsymbol{K}_{N_a} \\[1mm]
\boldsymbol{U}_1  &\cdots & \boldsymbol{U}_{N_a} \\
  \end{array}
\right], \boldsymbol{\mathbb{B}}=
\left[                 
  \begin{array}{c}   
   \ \,\, \boldsymbol{G}_1^+  \\[0.5mm]
   \ \,\, \vdots \\[0.5mm]
   \ \,\, \boldsymbol{G}_{N_a}^+  \\[0.5mm]
   - \boldsymbol{G}_1^-  \\[0.5mm]
   \ \,\, \vdots \\[0.5mm]
   - \boldsymbol{G}_{N_a}^-  \\[0.5mm]
   \ \,\, \boldsymbol{R}_1^+  \\[0.5mm]
   \ \,\, \vdots \\[0.5mm]
   \ \,\, \boldsymbol{R}_{N_a}^+  \\[0.5mm]
   - \boldsymbol{R}_1^-  \\[0.5mm]
   \ \,\, \vdots \\[0.5mm]
   - \boldsymbol{R}_{N_a}^-  \\[0.5mm]
	\boldsymbol{\Upsilon} \\[0.5mm]
   \boldsymbol{\Delta}  \\[0.5mm]
  \end{array}
  \right] \notag \\[-4mm]
  & \quad\quad\quad\underbrace{\quad\quad}_{\Large{\boldsymbol{\mathbb{A}}_1}} \quad \quad\ \quad\underbrace{\quad\quad}_{\Large{\boldsymbol{\mathbb{A}}_{N_a}}} \notag 
\end{align}

\subsection{Confidentiality Issues}

Ideally, to formulate $(\rm P_0)$, each regional ISO or a centralized entity should first form $\boldsymbol{Q} $, $\boldsymbol{c}$, $\boldsymbol{\mathbb{A}}$, and $\boldsymbol{\mathbb{B}}$. However, formulating these matrices or vectors requires knowledge of system data of the entire grid. For example, if computations are done by regional ISOs, then each of them has to collect the generation cost information of generators in other areas to form $\boldsymbol{Q}$ and $\boldsymbol{c}$, the capacities and ramp rates of generators in other regions to build $\boldsymbol{\mathbb{B}}$, the system topologies and line parameters of other regions to form $\boldsymbol{U}_n$~($\forall n$) in $\boldsymbol{\mathbb{A}}$. Such information may however be confidential.

Besides, once regional ISOs solve $(\rm P_0)$, they would infer additional information from the solution, which includes the dispatchable generation of each area. This information, however, might be confidential as well. 

% Assuming that regional ISOs formulated and solved $(\rm P_0)$, they would also infer additional information from the solution, which is the dispatchable generation of each area. This information, however, might  be confidential as well. 

% Accordingly, the purpose of this paper is to enable ISOs formulate and solve $(\rm P_0)$

% the solution of $(\rm{P0})$ are actually the dispatchable generations of all areas, which are confidential as well. If $(\rm{P0})$ is formulated and solved by regional ISOs, the solution will also reveal the confidential information of other regions.

% Overall, both formulating and solving ($\rm P0$) will cause privacy issues among regional grids. 

\section{TE Technique with Unsolved Challenges}
To tackle the confidentiality issues mentioned in the previous section, the TE technique is adopted in this paper. This technique can effectively build and solve linear programming problems while protecting the confidential information of participants from exposure \cite{li2013privacyv,li2013privacyh,8031400}. However, there are still some unsolved challenges standing in the way of using this technique to solve CC-OPF problems (e.g., CC-OPF problems are not linear programming). Hence, this section first briefly revisits the TE technique assuming a linear programming model. After that, the unsolved challenges when using this method to solve CC-OPF problems are discussed.

% in the following the TE technique will be briefly introduced  this section first briefly revisit this method and then explain the challenges when using TBA to solve ($\rm P0$).

\subsection{TE Technique}
Ignoring the quadratic term in ($\rm P0$) leads to a linear programming problem $(\rm{L0})$:
\begin{alignat}{2}
  (\rm L0)\ \ \min_{\boldsymbol{x}} \quad &  \mathcal{L}_0 = \boldsymbol{c}^\top\!\boldsymbol{x}   \\
  \mbox{s.t.}\quad
  &\boldsymbol{\mathbb{A}} \boldsymbol{x}\leq \boldsymbol{\mathbb{B}}
\end{alignat}

The basic idea of the TE technique is to utilize random matrices to encrypt the decision variable (i.e., dispatchable generation) of each region, which, in the meantime, also masks the confidential information of this area. Specifically, regional ISO $n$ first randomly generates an invertible matrix $\boldsymbol{M}_n \in \mathfrak{R} ^{H_n \times H_n}$. This matrix, called the encryption matrix, is only known by ISO $n$. Then, ISO $n$ transforms its own decision variable $\boldsymbol{x}_n$ into the encrypted decision variable $\boldsymbol{\bar{x}}_n \in \mathfrak{R}^{H_n \times 1}$ using $\boldsymbol{M}_n$. The relationship between $\boldsymbol{x}_n$ and $\boldsymbol{\bar{x}}_n$ is
\begin{equation}
	\boldsymbol{x}_n = \boldsymbol{M}_n\boldsymbol{\bar{x}}_n \label{mask n} 
\end{equation}
If we define
\begin{align}
	\boldsymbol{M} & = 
\left[                 
  \begin{array}{ccc}   
    \boldsymbol{M}_1  & \cdots & 0 \\
    \vdots  &   \ddots & \vdots \\[1mm]
    0 &   \cdots & \boldsymbol{M}_{N_a} \\
  \end{array}
\right] \in \mathfrak{R} ^{H\times H} \notag
\end{align}
then the following equation holds: 
\begin{equation}
	\boldsymbol{x} = \boldsymbol{M}\boldsymbol{\bar{x}} \label{eq:mask} 
\end{equation}
Substituting \eqref{eq:mask} into $(\rm L0)$ yields:
\begin{alignat}{2}
  \min_{\boldsymbol{\bar{x}}} \quad & \overline{\mathcal{L}} =  \left[\boldsymbol{c}_1\boldsymbol{M}_1 \ \  \cdots \ \  \boldsymbol{c}_{N_a}\boldsymbol{M}_{N_a} \right]^\top\!\boldsymbol{\bar{x}}  \notag  \\
 \mbox{s.t.}\quad
 &\left[\boldsymbol{\mathbb{A}}_1\boldsymbol{M}_1 \ \ \cdots  \ \  \boldsymbol{\mathbb{A}}_{N_a}\boldsymbol{M}_{N_a} \right]^\top\!\boldsymbol{\bar{x}} \leq \boldsymbol{\mathbb{B}} \notag
 \end{alignat}
Noticeably, after the transformation, $\boldsymbol{x}_n$ becomes $\boldsymbol{\bar{x}}_n$. This means that, once the transformed problem is solved, only ISO $n$ can recover $\boldsymbol{x}_n$ from $\boldsymbol{\bar{x}}_n$ using \eqref{mask n}. 
% As a result, the confidential information about the dispatchable power generation of each region is preserved. 
Also, the generation cost $\boldsymbol{c}_n$ is encrypted into $\boldsymbol{c}_n\boldsymbol{M}_n$ while $\boldsymbol{\mathbb{A}}_n$ is encrypted into $\boldsymbol{\mathbb{A}}_n\boldsymbol{M}_n$ after the transformation. Consequently, ISO $n$ can freely share $\boldsymbol{c}_n\boldsymbol{M}_n$ and $\boldsymbol{\mathbb{A}}_n\boldsymbol{M}_n$ with others, as no one can deduce $\boldsymbol{c}_n$ or $\boldsymbol{\mathbb{A}}_n$ from the exchanged information. Overall, $\boldsymbol{x}_n$, $\boldsymbol{c}_n$, and $\boldsymbol{\mathbb{A}}_n$ in $(\rm{L0})$ can be effectively masked by the encryption matrix $\boldsymbol{M}_n$. 

% , except for $\boldsymbol{\mathbb{B}}$. 

% As for $\boldsymbol{\mathbb{B}}$

However, $\boldsymbol{M}_n$ cannot obscure the information in $\boldsymbol{\mathbb{B}}$. But the way to protect each ISO's confidential data in $\boldsymbol{\mathbb{B}}$ is straightforward. For instance, to protect $\boldsymbol{G}_n^+$ in $\boldsymbol{\mathbb{B}}$ from disclosure, ISO $n$ can equivalently modify the corresponding inequality constraint from $\boldsymbol{J}_n\boldsymbol{x}_n \leq \boldsymbol{G}^{+}_n$ to:
\begin{align}
	\boldsymbol{\bar{J}}_n\boldsymbol{x}_n \leq \boldsymbol{\bar{G}}^{+}_n \label{eq:case_gen_upper_constraint_one_mask}
\end{align}
where $\boldsymbol{\bar{J}}_n = \boldsymbol{\varphi}_n\boldsymbol{J}_n$ and $\boldsymbol{\bar{G}}^{+}_n = \boldsymbol{\varphi}_n\boldsymbol{G}^{+}_n$. Matrix $\boldsymbol{\varphi}_n \in \mathfrak{R} ^{H_n \times H_n } $, randomly generated and privately hold by ISO $n$, is a diagonal matrix whose non-zero elements are positive. After introducing $\boldsymbol{M}_n$, \eqref{eq:case_gen_upper_constraint_one_mask} further becomes:
\begin{align}
	\boldsymbol{\bar{J}}_n\boldsymbol{M}_n\boldsymbol{\bar{x}}_n  \leq \boldsymbol{\bar{G}}^{+}_n  \notag
\end{align}
As a result, the capacity information of generators in region $n$ is encrypted into $\boldsymbol{\bar{G}}^{+}_n$. Similar operations can be done to $\boldsymbol{G}^{-}_n$, $\boldsymbol{R}^+_n$, and $\boldsymbol{R}^-_n$ to encrypt them into $\boldsymbol{\bar{G}}^{-}_n$, $\boldsymbol{\bar{R}}^+_n$, and $\boldsymbol{\bar{R}}^-_n$, respectively. Simultaneously, $\boldsymbol{J}_n$ and $\boldsymbol{E}_n$, i.e., the coefficients in the corresponding inequality constraints, are also modified to $\boldsymbol{\bar{J}}_n$ and $\boldsymbol{\bar{E}}_n$, respectively. For distinction, this paper uses $\boldsymbol{\bar{\mathbb{A}}}$ and $\boldsymbol{\bar{\mathbb{B}}}$ to separately denote $\boldsymbol{\mathbb{A}}$ and $\boldsymbol{\mathbb{B}}$ after the above modifications.

Finally, the original linear problem $(\rm L0)$ is transformed into its encrypted but equivalent version $(\rm L1)$:
\begin{alignat}{2}
(\rm L1)\ \  \min_{\boldsymbol{\bar{x}}} \quad & \mathcal{L}_1 =   \left[\boldsymbol{c}_1\boldsymbol{M}_1 \cdots \boldsymbol{c}_{N_a}\boldsymbol{M}_{N_a} \right]^\top\!\boldsymbol{\bar{x}}  \notag  \\
\mbox{s.t.}\quad
&\left[\boldsymbol{\bar{\mathbb{A}}}_1\boldsymbol{M}_1 \cdots \boldsymbol{\bar{\mathbb{A}}}_{N_a}\boldsymbol{M}_{N_a} \right]^\top\!\boldsymbol{\bar{x}} \leq \boldsymbol{\bar{\mathbb{B}}} \notag
\end{alignat}
Consequently, ISO $n$ can share its encrypted information, including $\boldsymbol{c}_n\boldsymbol{M}_n$, $\boldsymbol{\bar{\mathbb{A}}}_n\boldsymbol{M}_n$, $\boldsymbol{\bar{G}}^+_n$, $\boldsymbol{\bar{G}}^-_n$, $\boldsymbol{\bar{R}}^+_n$, and $\boldsymbol{\bar{R}}^-_n$, with other ISOs to allow each to build $(\rm L1)$. Once $(\rm L1)$ is solved, only ISO $n$ can derive its desired optimal solution, $\boldsymbol{x}_n^*$, from the encrypted optimal solution $\boldsymbol{\bar{x}}^*$. Clearly, no confidential data is leaked.

% First, both $\boldsymbol{\Upsilon}$ and $\boldsymbol{\Delta}$ are aggregations of loads, quantiles, and line capacities. That is to say, sharing $\boldsymbol{\Upsilon}$ and $\boldsymbol{\Delta}$ will not expose private information of individual area. Second, although areas refuse to directly share their generator capacities and ramp rates with others, they can instead share the masked but effective information. For example, the upper capacity constraints for generators in area $n$ is 

\subsection{Unsolved Challenges}

Although the TE technique is a confidentiality-preserving optimization method without parameter tunings, this method cannot be directly applied to the CC-OPF problem, i.e., $(\rm P_0)$. More specifically, there are three unsolved challenges standing in the way:
\begin{enumerate}
	\item As mentioned earlier, the TE technique is only proven to be effective for linear programming so far, but $(\rm P0)$ is a quadratic programming problem. Hence, the first challenge is to prove the method's effective application to quadratic problems.
	\item To use the TE technique, ISO $n$ must know $\boldsymbol{\bar{\mathbb{A}}}_n$. In $(\rm P0)$, however, ISO $n$ is unable to generate the complete $\boldsymbol{\bar{\mathbb{A}}}_n$ independently, because the sub matrix $\boldsymbol{U}_n$ in $\boldsymbol{\bar{\mathbb{A}}}_n$, known as the coupled coefficient matrix among dispatchable generations across regions, consists of the mapping coefficients that map nodal power injections to all the system states. These mapping coefficients are not locally obtainable unless global information, including the system topology and line parameters, is known. ISO $n$ does not have the global information, making $\boldsymbol{U}_n$ unobtainable and thereby the TE technique unusable. Given this, how to enable ISO $n$ to obtain $\boldsymbol{U}_n$ under the protection of regions' confidentiality, is the second unsolved challenge.
	\item Both $\boldsymbol{\Upsilon}$ and $\boldsymbol{\Delta}$ in $\boldsymbol{\bar{\mathbb{B}}}$ are aggregations of the regions' confidential information, including load data, line parameters, etc. Hence, the third challenge is to allow ISO $n$ to acquire $\boldsymbol{\Upsilon}$ and $\boldsymbol{\Delta}$ without having access to other ISOs' confidential data.
\end{enumerate}

The next section will address these challenges and provide solutions to each of them.

\section{Solutions to Unsolved Challenges}
In response to the three unsolved challenges mentioned above, this section first proves the TE technique's adaptability to quadratic programming. Then, a fast distributed algorithm for solving linear equation systems in a confidentiality-preserving manner is proposed, for the aim of helping ISO $n$ obtain $\boldsymbol{U}_n$. Finally, this section provides a distributed way to allow ISO $n$ to acquire $\boldsymbol{\Upsilon}$ and $\boldsymbol{\Delta}$ without any need for confidential information.

\subsection{Adaptability Proof}

After introducing the encryptions mentioned in \eqref{eq:mask} and \eqref{eq:case_gen_upper_constraint_one_mask}, the original CC-OPF problem $(\rm P0)$, a typical quadratic programming problem, is transformed into the encrypted version $(\rm P1)$:
\begin{alignat}{2}
  (\rm P1)\ \ \min_{\boldsymbol{\bar{x}}} \quad & \mathcal{F}_1 =  \frac{1}{2}\boldsymbol{\bar{x}}^\top \! \boldsymbol{\bar{Q}}\boldsymbol{\bar{x}} + \boldsymbol{c}^\top\!\boldsymbol{M}\boldsymbol{\bar{x}} \notag \\
  \mbox{s.t.}\quad
  &\boldsymbol{\bar{\mathbb{A}}} \boldsymbol{M}\boldsymbol{\bar{x}}\leq \boldsymbol{\bar{\mathbb{B}}} \notag
  \end{alignat}
  where
  \begin{equation}
    \boldsymbol{\bar{Q}}=\boldsymbol{M} ^\top \! \boldsymbol{Q} \boldsymbol{M} =  
    \left[                 
      \begin{array}{ccc}   
      \boldsymbol{M}_1^\top \boldsymbol{\Lambda}_1 \boldsymbol{M}_1 & \cdots & 0 \\[1mm]
      0 &   \ddots & 0 \\[1mm]
      0 & \cdots & \boldsymbol{M}_{N_a}^\top \boldsymbol{\Lambda}_{N_a} \boldsymbol{M}_{N_a} \\
      \end{array}
    \right] \notag
\end{equation}

Proving the TE technique's adaptability to $(\rm P0)$, is equivalent to proving the following two propositions. 
% \begin{enumerate}
% 	\item \textbf{Existence of global optimal solution}, i.e., $(\rm P1)$ has an unique global optimal solution.
% 	\item \textbf{Identity of global optimal solutions}, i.e., if $\boldsymbol{\bar{x}}^*$ is the global optimal solution of ($\rm P1$), then $\boldsymbol{x}^*=\boldsymbol{M}\boldsymbol{\bar{x}}^*$ is the global optimal solution of ($\rm P0$), and vice versa.
% \end{enumerate}

% The following will prove these two properties.

% \vspace{0.2cm}
\noindent \textbf{Proposition 1.} There exists a global optimal solution of ($\rm P1$).
% \vspace{0.2cm}

\noindent \textit{Proof}. As described earlier, $\boldsymbol{Q}$ is a diagonal matrix whose non-zero elements are all positive. Hence, $\boldsymbol{Q}$ is a positive definite matrix. According to the definition of being positive definite, the following equation holds:
\begin{equation}
  \boldsymbol{x}^{\top}\boldsymbol{Q}\boldsymbol{x}>0, \ \forall \boldsymbol{x} \neq \boldsymbol{0} \notag
\end{equation}
Since 
\begin{equation}
  \boldsymbol{x}^{\top}\boldsymbol{Q}\boldsymbol{x} = \boldsymbol{\bar{x}}^\top \! \boldsymbol{M} ^\top \!\boldsymbol{Q} \boldsymbol{M}\boldsymbol{\bar{x}} = \boldsymbol{\bar{x}}^\top \boldsymbol{\bar{Q}}\boldsymbol{\bar{x}} \notag
\end{equation}
it follows that 
\begin{align}
  \boldsymbol{\bar{x}}^\top \boldsymbol{\bar{Q}}\boldsymbol{\bar{x}} >0, \ \forall \boldsymbol{x} \neq \boldsymbol{0}  \label{eq:Q bar}
\end{align}
Note that $\boldsymbol{Q}$ is also a real symmetric matrix, hence 
\begin{equation}
  \boldsymbol{\bar{Q}}^\top\! = \boldsymbol{M} ^\top \! \boldsymbol{Q}^\top \boldsymbol{M} = \boldsymbol{M} ^\top \! \boldsymbol{Q} \boldsymbol{M} = \boldsymbol{\bar{Q}}  \notag
\end{equation}
holds, i.e., $\boldsymbol{\bar{Q}}$ is a real symmetric matrix as well. 

As $\boldsymbol{M}_n$ is invertible, we can form
\begin{align}
  \boldsymbol{M}^{-1} & = 
\left[                 
\begin{array}{ccc}   
  \boldsymbol{M}_1^{-1}  & \cdots & 0 \\
  \vdots &   \ddots & 0 \\[1mm]
  0 &   \cdots & \boldsymbol{M}_{N_a}^{-1} \\
\end{array}
\right] \notag
\end{align}
meaning that $\boldsymbol{M}$ is also invertible. Accordingly, $\rm rank(\boldsymbol{M}^{-1})= \textit{H}$ holds, where $\rm rank(\cdot)$ denotes the rank function. Therefore, in the space of $\mathfrak{R} ^{H \times 1 }$, matrix $\boldsymbol{M}^{-1}$ fully maps $\forall \boldsymbol{x} \neq \boldsymbol{0} $ to $\forall \boldsymbol{\bar{x}} \neq \boldsymbol{0}$ using the relationship $\boldsymbol{\bar{x}}  = \boldsymbol{M}^{-1}\boldsymbol{x}$. That is to say, $\forall \boldsymbol{x} \neq \boldsymbol{0} $ yields $\forall \boldsymbol{\bar{x}} \neq \boldsymbol{0}$. As a result, \eqref{eq:Q bar} is equivalent to 
\begin{align}
  \boldsymbol{\bar{x}}^\top \boldsymbol{\bar{Q}}\boldsymbol{\bar{x}} >0, \forall \boldsymbol{\bar{x}} \neq \boldsymbol{0}  \label{eq:Q bar 2}
\end{align}

To summarize, $\boldsymbol{\bar{Q}}$ is a real symmetric matrix and \eqref{eq:Q bar 2} holds. Then, according to the definition of being positive definite, $\boldsymbol{\bar{Q}}$ is a positive definite matrix. Consequently, ($\rm P1$) is strictly convex, thereby ensuring the existence of a unique global optimal solution.$\hfill\blacksquare$
 
% \vspace{0.2cm}
\noindent \textbf{Proposition 2.} If $\boldsymbol{\bar{x}}^*$ is the global optimal solution of ($\rm P1$), then $\boldsymbol{x}^*=\boldsymbol{M}\boldsymbol{\bar{x}}^*$ is the global optimal solution of ($\rm P0$), and vice versa.
% \vspace{0.2cm}

\noindent \textit{Proof}. We start by introducing the semi-encrypted problem ($\rm P2$) as follows:
\begin{alignat}{2}
  (\rm P2)\ \ \min_{\boldsymbol{x}} \quad &  \mathcal{F}_0 =  \frac{1}{2}\boldsymbol{x}^\top \! \boldsymbol{Q} \boldsymbol{x} + \boldsymbol{c}^\top\!\boldsymbol{x} \notag \\
  \mbox{s.t.}\quad
  &\boldsymbol{\bar{\mathbb{A}}} \boldsymbol{x}\leq \boldsymbol{\bar{\mathbb{B}}} \notag
\end{alignat}
Compared to ($\rm P0$), ($\rm P2$) is formulated with the equivalently-modified $\boldsymbol{\bar{\mathbb{A}}}$ and $\boldsymbol{\bar{\mathbb{B}}}$. In other words, ($\rm P0$) and ($\rm P2$) are equivalent.

If $\boldsymbol{\bar{x}}^*$ is the global optimal solution of ($\rm P1$), then $\boldsymbol{\bar{x}}^*$ satisfies the Karush-Kuhn-Tucker (KKT) condition of ($\rm P1$):
\begin{equation}
\label{eq:KKT P1}
\left\{
\begin{split}
& \boldsymbol{M} ^\top \!( \boldsymbol{Q} \boldsymbol{M}\boldsymbol{\bar{x}}^* + \boldsymbol{c} + \boldsymbol{\bar{\mathbb{A}}}^\top \!\boldsymbol{\boldsymbol{\lambda}_L}) = 0 \\
& \boldsymbol{\bar{\mathbb{A}}}\boldsymbol{M}\boldsymbol{\bar{x}}^*\leq \boldsymbol{\bar{\mathbb{B}}}\\
& \boldsymbol{\boldsymbol{\lambda}_L} \geq 0 \\
& \boldsymbol{\boldsymbol{\lambda}_L}^\top\!\boldsymbol{\bar{\mathbb{A}}}\boldsymbol{M}\boldsymbol{\bar{x}}^* - \boldsymbol{\boldsymbol{\lambda}_L}^\top\boldsymbol{\bar{\mathbb{B}}}=0
\end{split} \right.
\end{equation}
where $\boldsymbol{\lambda}_L$ is the vector of lagrangian multipliers. Note that $\boldsymbol{M}$ is invertible, hence $\boldsymbol{M} ^\top \!( \boldsymbol{Q} \boldsymbol{M}\boldsymbol{\bar{x}}^* + \boldsymbol{c} + \boldsymbol{\bar{\mathbb{A}}}^\top \!\boldsymbol{\boldsymbol{\lambda}_L}) = 0 $ in \eqref{eq:KKT P1} is equivalent to:
\begin{equation}
\boldsymbol{Q} \boldsymbol{M}\boldsymbol{\bar{x}}^* + \boldsymbol{c} + \boldsymbol{\bar{\mathbb{A}}}^\top \!\boldsymbol{\boldsymbol{\lambda}_L}=0 \label{eq:essence}
\end{equation}
Substituting $\boldsymbol{\bar{x}}^* =\boldsymbol{M}^{-1}\boldsymbol{x}^*$ into \eqref{eq:KKT P1} and \eqref{eq:essence} leads to
\begin{equation}
\label{eq:KKT P0}
\left\{
\begin{split}
&  \boldsymbol{Q}\boldsymbol{x}^* + \boldsymbol{c} + \boldsymbol{\bar{\mathbb{A}}}^\top \!\boldsymbol{\boldsymbol{\lambda}_L} = 0 \\
& \boldsymbol{\bar{\mathbb{A}}}\boldsymbol{x}^*\leq \boldsymbol{\bar{\mathbb{B}}}\\
& \boldsymbol{\boldsymbol{\lambda}_L} \geq 0 \\
& \boldsymbol{\boldsymbol{\lambda}_L}^\top\!\boldsymbol{\bar{\mathbb{A}}}\boldsymbol{x}^* - \boldsymbol{\boldsymbol{\lambda}_L}^\top\boldsymbol{\bar{\mathbb{B}}}=0
\end{split} \right.
\end{equation}
Noticeably, \eqref{eq:KKT P0} are exactly the KKT conditions of ($\rm P2$). Most importantly, $\boldsymbol{x}^*$ satisfies this KKT conditions, thereby making $\boldsymbol{x}^*$ the global optimal solution of ($\rm P2$).

Conversely, if $\boldsymbol{x}^*$ is the global optimal solution of ($\rm P2$), then $\boldsymbol{x}^*$ meets the KKT conditions of ($\rm P2$), i.e., \eqref{eq:KKT P0}. Substituting $\boldsymbol{x}^*=\boldsymbol{M}\boldsymbol{\bar{x}}^*$ into \eqref{eq:KKT P0} generates:
\vspace{-3pt}
\begin{equation}
\label{eq:KKT P11}
\left\{
\begin{split}
& \boldsymbol{Q} \boldsymbol{M}\boldsymbol{\bar{x}}^* + \boldsymbol{c} + \boldsymbol{\bar{\mathbb{A}}}^\top \!\boldsymbol{\boldsymbol{\lambda}_L} = 0 \\
& \boldsymbol{\bar{\mathbb{A}}}\boldsymbol{M}\boldsymbol{\bar{x}}^*\leq \boldsymbol{\bar{\mathbb{B}}}\\
& \boldsymbol{\boldsymbol{\lambda}_L} \geq 0 \\
& \boldsymbol{\boldsymbol{\lambda}_L}^\top\!\boldsymbol{\bar{\mathbb{A}}}\boldsymbol{M}\boldsymbol{\bar{x}}^* - \boldsymbol{\boldsymbol{\lambda}_L}^\top\boldsymbol{\bar{\mathbb{B}}}=0
\end{split} \right.
\end{equation}
Since $\boldsymbol{M}$ is invertible, multiplying the left and right sides of the first equation in \eqref{eq:KKT P11} by $\boldsymbol{M}$ yields
\begin{equation}
\boldsymbol{M} ^\top \!( \boldsymbol{Q} \boldsymbol{M}\boldsymbol{\bar{x}}^* + \boldsymbol{c} + \boldsymbol{\bar{\mathbb{A}}}^\top \!\boldsymbol{\boldsymbol{\lambda}_L}) = 0 \label{eq:essence2}
\end{equation}
Replacing the first equation in \eqref{eq:KKT P11} with \eqref{eq:essence2} results in the KKT condition of ($\rm P1$), i.e., \eqref{eq:KKT P1}. Clearly, $\boldsymbol{\bar{x}}^*$ fulfills these KKT conditions, thus ensuring $\boldsymbol{\bar{x}}^*$ to be the global optimal solution of ($\rm P1$). 

The above proves that if $\boldsymbol{\bar{x}}^*$ is the global optimal solution of ($\rm P1$), then $\boldsymbol{x}^*=\boldsymbol{M}\boldsymbol{\bar{x}}^*$ is the global optimal solution of ($\rm P2$), and vice versa. Note that ($\rm P0$) and ($\rm P2$) are equivalent. Consequently, Proposition 2 holds.  $\hfill\blacksquare$ 

% Finally, the following proposition holds in terms of the proved equivalent proposition: if $\boldsymbol{\bar{x}}^*$ is the global optimal solution of ($\rm P1$), then $\boldsymbol{x}^*=\boldsymbol{M}\boldsymbol{\bar{x}}^*$ is the global optimal solution of ($\rm P0$), and vice versa. $\hfill\blacksquare$ 

% \vspace{0.2cm}

The above two propositions guarantee the TE technique's adaptability to quadratic programming. 

% Thus, each ISO only needs to build and solve ($\rm P1$) instead ($\rm P0$).

\subsection{Coupled Coefficient Matrix Calculation}

According to \eqref{eq:Un}, the coupled coefficient matrix $\boldsymbol{U}_n$ is actually formed by $\boldsymbol{\Phi}_n$. The latter, defined by \eqref{eq:phin}, consists of the mapping coefficients that map the dispatchable generations in region $n$ to the system states of the whole grid. Therefore, calculating $\boldsymbol{U}_n$ corresponds to determining the mapping coefficients in $\boldsymbol{\Phi}_n$. 

To this end, the DLPF model of the whole grid is needed. This model is given by
\begin{align}
  \label{eq:DLPF_multi_region} 
  \left[                 
    \begin{array}{c}   
      \boldsymbol{G}_{Q,1} \\     \boldsymbol{G}_{P,1} \\ \vdots \\\boldsymbol{G}_{Q,{N_a}} \\ \boldsymbol{G}_{P,{N_a}}\\
    \end{array}
  \right] 
  \left[                 
    \begin{array}{c}   
      \boldsymbol{V}_{1} \\ \boldsymbol{\theta}_{1} \\ \vdots \\   \boldsymbol{V}_{{N_a}} \\  \boldsymbol{\theta}_{{N_a}} \\
    \end{array}
  \right] = 
  \left[                 
    \begin{array}{c}   
      \boldsymbol{\widetilde{Q}}_{1} \\ \boldsymbol{\widetilde{P}}_{1} \\ \vdots \\   \boldsymbol{\widetilde{Q}}_{N_a} \\  \boldsymbol{\widetilde{P}}_{N_a} \\
    \end{array}
  \right] 
\end{align}
where $\boldsymbol{V}_n$ consists of the voltage magnitudes of the $PQ$ nodes in region $n$, while $\boldsymbol{\widetilde{Q}}_n$ is formed by the reactive power injections at the same nodes. Besides, $\boldsymbol{\theta}_n$ is composed of the voltage angles of the $PQ$ and $PV$ nodes in region $n$, while  $\boldsymbol{\widetilde{P}}_n$ includes the active power injections at the same nodes. In addition, $\boldsymbol{G}_{Q,n}$ denotes the line conductances between the $PQ$ nodes in region $n$ and all the other nodes in the grid, while $\boldsymbol{G}_{P,n}$ includes the corresponding line susceptances. Apparently, ISO $n$ only knows $\boldsymbol{G}_{Q,n}$ and $\boldsymbol{G}_{P,n}$ in \eqref{eq:DLPF_multi_region}, as the related lines are either the inner lines within region $n$ or the tie lines between region $n$ and other regions. 

% Other line parameters are confidentiality of other regions, i.e., secrets to ISO $n$.

We further define $\boldsymbol{G}$ as:
\begin{align}
	\boldsymbol{G} = 
	\left[                 
	  \begin{array}{ccccc}   
		\boldsymbol{G}_{Q,1}^\top &     \boldsymbol{G}_{P,1}^\top & \cdots &     \boldsymbol{G}_{Q,{N_a}}^\top &     \boldsymbol{G}_{P,{N_a}}^\top
	  \end{array}
	\right]^\top \notag
	\end{align}
Then 
% \begin{align}
% 	\left[                 
% 	  \begin{array}{ccccc}   
% 		\boldsymbol{V}_{1}^\top &   \boldsymbol{\theta}_{1}^\top & \vdots &   \boldsymbol{V}_{{N_a}}^\top &  \boldsymbol{\theta}_{{N_a}}^\top \\
% 	  \end{array}
% 	\right]  = 
% 	\boldsymbol{G}^{-1}
% 	\left[                 
% 	  \begin{array}{ccccc}   
% 		\boldsymbol{\widetilde{Q}}_{1}^\top & \boldsymbol{\widetilde{P}}_{1}^\top & \vdots &   \boldsymbol{\widetilde{Q}}_{{N_a}}^\top &  \boldsymbol{\widetilde{P}}_{{N_a}}^\top \\
% 	  \end{array}
% 	\right]  \notag
% \end{align}
\begin{align}
	\left[                 
		\boldsymbol{V}_{1}^\top ,   \boldsymbol{\theta}_{1}^\top,   \cdots,    \boldsymbol{V}_{{N_a}}^\top,   \boldsymbol{\theta}_{{N_a}}^\top 
	\right]^\top \!\!\!  = \! 
	\boldsymbol{G}^{-1} \!
	\left[                 
		\boldsymbol{\widetilde{Q}}_{1}^\top , \boldsymbol{\widetilde{P}}_{1}^\top,  \cdots,     \boldsymbol{\widetilde{Q}}_{{N_a}}^\top,  \boldsymbol{\widetilde{P}}_{{N_a}}^\top 
	\right]^\top  \notag
\end{align}
% \begin{align}
% 	\left[                 
% 	  \begin{array}{c}   
% 		\boldsymbol{V}_{1} \\ \boldsymbol{\theta}_{1} \\ \vdots \\   \boldsymbol{V}_{{N_a}} \\  \boldsymbol{\theta}_{{N_a}} \\
% 	  \end{array}
% 	\right]  & = 
% 	\boldsymbol{G}^{-1}
% 	\left[                 
% 	  \begin{array}{c}   
% 		\boldsymbol{\widetilde{Q}}_{1} \\ \boldsymbol{\widetilde{P}}_{1} \\ \vdots \\   \boldsymbol{\widetilde{Q}}_{{N_a}} \\  \boldsymbol{\widetilde{P}}_{{N_a}} \\
% 	  \end{array}
% 	\right]  \notag
% \end{align}
where $\boldsymbol{G}^{-1}$ is denoted by
\begin{align}
	\boldsymbol{G}^{-1} = \left[                 
	  \begin{array}{ccccc}   
		\boldsymbol{B}_{Q,1}^\top &     \boldsymbol{B}_{P,1}^\top & \cdots &     \boldsymbol{B}_{Q,{N_a}}^\top &     \boldsymbol{B}_{P,{N_a}}^\top
	  \end{array}
	\right] \notag
\end{align}

Clearly, once ISO $n$ obtains $\boldsymbol{B}_{Q,n}$ and $\boldsymbol{B}_{P,n}$, it can directly form the desired matrix $\boldsymbol{\Phi}_n$. Computing $\boldsymbol{B}_{Q,n}$ and $\boldsymbol{B}_{P,n}$ requires the inverse calculation of $\boldsymbol{G}$. However, ISO $n$ only has partial information of $\boldsymbol{G}$, preventing ISO $n$ from calculating the inverse of $\boldsymbol{G}$. Actually, calculating the inverse of $\boldsymbol{G}$ can be considered as solving the following system of linear equations:
\begin{equation}
	\label{eq:ZYI}
	\left[                 
	  \begin{array}{ccccc}   
		\boldsymbol{G}_{Q,1}^\top \!\!  &  \!\!   \boldsymbol{G}_{P,1}^\top \!\! &  \!\! \cdots \!\! &   \!\!  \boldsymbol{G}_{Q,{N_a}}^\top  \!\! & \!\!    \boldsymbol{G}_{P,{N_a}}^\top
	  \end{array}
	\right]\left[                 
		\begin{array}{c}   
		  \boldsymbol{B}_{Q,1} \\     \boldsymbol{B}_{P,1}\\ \vdots \\     \boldsymbol{B}_{Q,{N_a}} \\    \boldsymbol{B}_{P,{N_a}}
		\end{array}
	  \right] = \boldsymbol{I}
\end{equation}
where $\boldsymbol{I}$ is an identity matrix with the appropriate dimension, and $\boldsymbol{B}_{Q,1} \ \cdots  \boldsymbol{B}_{P,N_a}$ are the solution of this system. Note that the linear equation system in \eqref{eq:ZYI} has a special feature: ISO $n$ only knows some equations of the system, i.e., ISO $n$ only has $\boldsymbol{G}_{Q,n}$ and $\boldsymbol{G}_{P,n}$ at hand. This is a typical problem in the field of distributed computing, which has drawn much attention recently \cite{azizan2019distributed}. However, the state-of-the-art methods to solve this problem usually require a lot of iterations, leading to low computational efficiency. Therefore, we propose a fast distributed method that can solve \eqref{eq:ZYI} in a confidentiality-preserving way.

First, ISO $n$ randomly generates two invertible and private matrices: $\boldsymbol{W}_{Q,n}$ and $\boldsymbol{W}_{P,n}$. These two matrices are used to adapt $\boldsymbol{B}_{Q,n}$ and $\boldsymbol{B}_{P,n}$ into their encrypted versions, i.e.,  $\boldsymbol{\bar{B}}_{Q,n}$ and $\boldsymbol{\bar{B}}_{P,n}$:
\begin{align}
	\boldsymbol{B}_{Q,n}  = \boldsymbol{W}_{Q,n} \boldsymbol{\bar{B}}_{Q,n}, \ \  \boldsymbol{B}_{P,n}  = \boldsymbol{W}_{P,n} \boldsymbol{\bar{B}}_{P,n} \label{eq:mask_YQP} 
\end{align}
Further, we define $\boldsymbol{W}$ as:
\begin{align}
	\boldsymbol{W} & = 
\left[                 
  \begin{array}{ccccc}   
    \boldsymbol{W}_{Q,1} & 0 & \cdots & 0 & 0\\[1mm]
    0 & \boldsymbol{W}_{P,1} & \cdots & 0 & 0\\[1mm]
    0 & 0 &   \ddots & 0 & 0 \\[1mm]
    0 & 0 & \cdots & \boldsymbol{W}_{Q,{N_a}} & 0\\[1mm]
    0 & 0 & \cdots & 0 & \boldsymbol{W}_{P,{N_a}}\\
  \end{array}
\right] \notag
\end{align}
Then the following equation holds:
% \begin{align}
% 	\label{eq:mask_Y}
% 	\left[                 
% 	  \begin{array}{c}   
% 		\boldsymbol{B}_{Q,1} \\     \boldsymbol{B}_{P,1}\\ \vdots \\     \boldsymbol{B}_{Q,{N_a}} \\    \boldsymbol{B}_{P,{N_a}}
% 	  \end{array}
% 	\right] = \boldsymbol{W}
% 		\left[                 
% 	  \begin{array}{c}   
% 		\boldsymbol{\bar{B}}_{Q,1} \\     \boldsymbol{\bar{B}}_{P,1}\\ \vdots \\     \boldsymbol{\bar{B}}_{Q,{N_a}} \\    \boldsymbol{\bar{B}}_{P,{N_a}}
% 	  \end{array}
% 	\right]
% \end{align}
\begin{align}
	\left[                 
		\boldsymbol{B}_{Q,1}^\top ,   \cdots,     \boldsymbol{B}_{P,{N_a}}^\top 
	\right]^\top  = 
	\boldsymbol{W}
	\left[                 
		\boldsymbol{\bar{B}}_{Q,1}^\top ,  \cdots,     \boldsymbol{\bar{B}}_{P,{N_a}}^\top 
	\right]^\top  \label{eq:mask_Y}
\end{align}
Substituting \eqref{eq:mask_Y} into \eqref{eq:ZYI} generates
\begin{align}
	\label{eq:solve_Y}
	\left[                 
	  \begin{array}{ccc}   
		\boldsymbol{G}_{Q,1}^\top \boldsymbol{W}_{Q,1} \!\! & \!\! \cdots \!\! &  \!\!   \boldsymbol{G}_{P,{N_a}}^\top\boldsymbol{W}_{P,{N_a}}
	  \end{array}
	\right] \left[                 
	  \begin{array}{c}   
		\boldsymbol{\bar{B}}_{Q,1} \\ 
     \vdots \\   
    \boldsymbol{\bar{B}}_{P,{N_a}}
	  \end{array}
	\right] = \boldsymbol{I}
\end{align}

Since only ISO $n$ knows $\boldsymbol{G}_{Q,n}$, $\boldsymbol{G}_{P,n}$, $\boldsymbol{W}_{Q,n}$, and $\boldsymbol{W}_{Q,n}$, ISO $n$ can freely share the masked information $\boldsymbol{G}_{Q,n}^\top \boldsymbol{W}_{Q,n}$ and $\boldsymbol{G}_{P,n}^\top \boldsymbol{W}_{P,n}$ with other ISOs, as none of the other ISOs can deduce $\boldsymbol{G}_{Q,n}$ and $\boldsymbol{G}_{P,n}$ from the exchanged information. Once all the masked information is received from the other ISOs, each ISO can solve \eqref{eq:solve_Y} and obtain the encrypted solution. Eventually, ISO $n$ can recover $\boldsymbol{B}_{Q,n}$ and $\boldsymbol{B}_{P,n}$ from the encrypted solution using \eqref{eq:mask_YQP}. Consequently, the whole process is confidentiality-preserving.

It is worth mentioning that in the above process, only the sharing of masked information requires communications between ISOs --- the actual calculations can be implemented independently by each ISO. To make the whole process a distributed procedure, i.e. not all ISOs having to communicate with all other ISOs, an accelerated average consensus (AAC) algorithm \cite{aysal2008accelerated} is leveraged to adapt the information sharing process into a fully distributed calculation. 

The AAC algorithm is a graph-theory-based distributed method, where in the considered case ISOs are the vertices in the graph while communication lines between ISOs constitute the edges. During the iterations of this algorithm, each ISO only needs to communicate with its one-hop neighbors, yet the calculation result of every ISO will converge to the mean value of all ISOs' initial values. Hence, to realize the information sharing between ISOs by the AAC algorithm, ISO $n$ should first form its initial value $\boldsymbol{y}_n(0)$ using its encrypted information:
\begin{equation}
  \boldsymbol{y}_n(0) = 
	\begin{bmatrix}
		0 & \cdots & \boldsymbol{G}_{Q,n}^\top \boldsymbol{W}_{Q,n} & \boldsymbol{G}_{P,n}^\top \boldsymbol{W}_{P,n} & \cdots & 0 \notag
	\end{bmatrix}
\end{equation}
where all elements are zero except for the ($2n-1$)-th element $\boldsymbol{G}_{Q,n}^\top \boldsymbol{W}_{Q,n}$ and the $2n$-th element $\boldsymbol{G}_{P,n}^\top \boldsymbol{W}_{P,n}$. Then, ISO $n$'s initial value is updated according to:
\begin{align}
	\boldsymbol{y}_n(k+1) & = \beta \boldsymbol{y}_n^p(k+1) + (1-\beta) \boldsymbol{y}_n^w(k+1) \label{AAC algorithm} \\
	\boldsymbol{y}_n^w(k+1) & = \varpi_{n,n}\boldsymbol{y}_n(k) + \sum\nolimits_{m \in \boldsymbol{\Omega}_n } \varpi_{n,m}\boldsymbol{y}_m(k) \notag \\
	\boldsymbol{y}_n^p(k+1) & = \eta_1 \boldsymbol{y}_n^w(k+1) + \eta_2 \boldsymbol{y}_n(k) \notag
\end{align}
and $k$ is the iteration counter. All the parameters above, including $\beta$, $\varpi_{n,n}$, $\varpi_{n,m}$, $\eta_1$, and $\eta_2$, can be easily set using the generalized rules mentioned in \cite{aysal2008accelerated}. With this process, the iterative result $\boldsymbol{y}_n(k)$ of ISO $n$, will converge, i.e.,
\begin{equation}
	\lim\limits_{k \to \infty }{\boldsymbol{y}_n(k)} = \frac{1}{N_a} \sum\nolimits_{n=1}^{N_a} \boldsymbol{y}_n(0) \notag 
\end{equation}
Multiplying the converged result by $N_a$, ISO $n$ finally obtains
\begin{equation}
	\begin{bmatrix}
		\boldsymbol{G}_{Q,1}^\top \boldsymbol{W}_{Q,1} \!\! \!\! & \boldsymbol{G}_{P,1}^\top \boldsymbol{W}_{P,1} \!\! \!\! & \cdots \!\! \!\! & \boldsymbol{G}_{Q,{N_a}}^\top \boldsymbol{W}_{Q,{N_a}} \!\! \!\! & \boldsymbol{G}_{P,{N_a}}^\top \boldsymbol{W}_{P,{N_a}} \notag
	\end{bmatrix}
\end{equation} 
i.e., the masked information of all ISOs. 

Overall, the proposed fast distributed method for solving the linear equation system can be summarized as three steps: (i) ISO $n$ formulates and shares its masked information via the AAC algorithm; (ii) ISO $n$ solves \eqref{eq:solve_Y} and obtains the encrypted solution; (iii) ISO $n$ recovers $\boldsymbol{B}_{Q,n}$ and $\boldsymbol{B}_{P,n}$ from the encrypted solution using \eqref{eq:mask_YQP}. The proposed method is confidentiality-preserving, as the confidential information of ISO $n$ is strictly protected by the randomly-generated matrices $\boldsymbol{W}_{Q,n}$ and $\boldsymbol{W}_{Q,n}$. Also, this method has high efficiency, since only linear algebraic computations are required for encryption and decryption, not to mention that the AAC algorithm has a fast convergence rate given its accelerated design. 

Once ISO $n$ obtains $\boldsymbol{B}_{Q,n}$ and $\boldsymbol{B}_{P,n}$ by the proposed method, it can form $\boldsymbol{\Phi}_n$ according to \eqref{eq:phin} and build $\boldsymbol{U}_n$ based on \eqref{eq:Un}. 

\vspace{-0.2cm}
\subsection{Aggregation of Confidential Data}

Formulating $\boldsymbol{\Upsilon}$ and $\boldsymbol{\Delta}$ in $\boldsymbol{\bar{\mathbb{B}}}$ essentially corresponds to summing the confidential data (e.g., the load data) of different regions. Although the AAC algorithm allows ISOs to achieve this summation in a distributed manner (the mean value and the summation value are interchangeable), this algorithm will leak the exchanged data to ISOs, leading to the disclosure of confidential information. 

To enable ISOs to obtain the summation in both distributed and confidentiality-preserving manners, a privacy-preserving AAC (PP-AAC) algorithm proposed in \cite{9223745} is adopted. This algorithm introduces noise to mask the exchanged data between ISOs, thereby protecting their information. Most importantly, the PP-AAC algorithm can still guarantee the accuracy of the summation calculation.

Specifically, ISO $n$ first sets its confidential data, e.g., the load data, as its initial value $\boldsymbol{z}_n(0)$. Then, ISO $n$ initializes $\boldsymbol{y}_n(0)$ by 
\begin{equation}
		\boldsymbol{y}_n(0) = \boldsymbol{z}_n(0) + \delta_n(0)\notag
\end{equation}
where $\delta_n(0)$ is noise randomly selected from $[-\frac{\sigma}{2}\gamma, \frac{\sigma}{2}\gamma]$ by ISO $n$. Note that $\sigma>0$ and $\gamma \in [0,1)$. The iterative process using the AAC algorithm then starts. In the $k$-th iteration ($k\ge 1$) of the AAC algorithm, $\boldsymbol{y}_n(k)$ will be further masked by newly generated noise:
\begin{equation}
	\begin{split}
		& \boldsymbol{y}_n(k) \leftarrow \boldsymbol{y}_n(k) + \theta_n(k) \\
		& \theta_n(k) = \delta_n(k) - \delta_n(k-1) \notag
	\end{split}
\end{equation}
where $\delta_n(k)$ is randomly selected from $[-\frac{\sigma}{2}\gamma^{k+1}, \frac{\sigma}{2}\gamma^{k+1}]$ by ISO $n$. Finally, $\boldsymbol{y}_n(k)$ will converge to the mean of all ISOs' initial values, i.e., $\sum\nolimits_{n=1}^{N_a} \boldsymbol{z}_n(0)$, but the confidential information contained in $\boldsymbol{z}_n(0)$ is protected \cite{9223745}.  

To summarize, for aggregating confidential data, each ISO first sets its confidential information as the initial value of the PP-AAC algorithm and then participates in the iterative process. Once converged mean results are obtained, each ISO can acquire the desired summation (i.e., $\boldsymbol{\Upsilon}$ and $\boldsymbol{\Delta}$) with a simple multiplication.

\section{Distributed CC-OPF Method with Confidentiality Preservation}

Since the challenges when using the TE technique have been addressed in the previous section, this section summarizes and further elaborates the application of the proposed distributed CC-OPF method with confidentiality preservation based on the TE approach. 

% After that, the advantages of this method will be discussed.

% \subsection{Methodology}

The proposed confidentiality-preserving distributed CC-OPF method consists of five steps, namely formulating, encrypting, sharing, solving, and decrypting. Details are as follows:

\textbf{1) Formulating Step} ---  First, ISO $n$ forms $\boldsymbol{U}_n$ using the proposed fast distributed method for linear equation systems. Second, ISO $n$ forms $\boldsymbol{\Upsilon}$ and $\boldsymbol{\Delta}$ using the PP-AAC algorithm. Third, ISO $n$ formulates $\boldsymbol{\Lambda}_n$, $\boldsymbol{c}_n$, $\boldsymbol{\mathbb{A}}_n$, $\boldsymbol{G}_n^+$, $\boldsymbol{G}_n^-$, $\boldsymbol{R}_n^+$, and $\boldsymbol{R}_n^+$ independently. These computations can be carried out in parallel. 

\textbf{2) Encrypting Step} ---  First, ISO $n$ transforms $\boldsymbol{\mathbb{A}}_n$, $\boldsymbol{G}_n^+$, $\boldsymbol{G}_n^-$, $\boldsymbol{R}_n^+$, and $\boldsymbol{R}_n^+$ into $\boldsymbol{\bar{\mathbb{A}}}_n$, $\boldsymbol{\bar{G}}_n^+$, $\boldsymbol{\bar{G}}_n^-$, $\boldsymbol{\bar{R}}_n^+$,  and $\boldsymbol{\bar{R}}_n^+$. Second, ISO $n$ randomly generates encryption matrix $\boldsymbol{M}_n$ and uses it to encrypt $\boldsymbol{x}_n$ into $\boldsymbol{\bar{x}}_n$, simultaneously masking $ \boldsymbol{\Lambda}_n$, $\boldsymbol{c}_n$, and $\boldsymbol{\mathbb{A}}_n$ into $\boldsymbol{M}_n^\top \boldsymbol{\Lambda}_n\boldsymbol{M}_n$, $\boldsymbol{c}_n^\top\boldsymbol{M}_n$, and $\boldsymbol{\bar{\mathbb{A}}}_n\boldsymbol{M}_n$, respectively.

\textbf{3) Sharing Step} ---  ISO $n$ shares $\boldsymbol{\bar{G}}_n^+$, $\boldsymbol{\bar{G}}_n^-$, $\boldsymbol{\bar{R}}_n^+$, $\boldsymbol{\bar{R}}_n^+$, $\boldsymbol{\bar{\mathbb{A}}}_n\boldsymbol{M}_n$, $\boldsymbol{M}_n^\top \boldsymbol{\Lambda}_n\boldsymbol{M}_n$, and $\boldsymbol{c}_n^\top\boldsymbol{M}_n$ with other ISOs using the AAC algorithm.

\textbf{4) Solving Step} ---  ISO $n$ uses the received information to build and solve $(\rm P1)$, thus obtaining the encrypted optimal solution $\boldsymbol{\bar{x}}^*$.

\textbf{5) Decrypting Step} ---  ISO $n$ recovers its desired optimal solution $\boldsymbol{x}_n^*$ from $\boldsymbol{\bar{x}}^*$ using its encryption matrix $\boldsymbol{M}_n$.

The above five steps constitute the concise version of the proposed method. It should be noted that this version can only handle chance constraints with respect to nodal voltages included in the OPF problem. This is because $\boldsymbol{U}_n$ is composed of the elements in $\boldsymbol{B}_{Q,n}$ and $\boldsymbol{B}_{P,n}$, and the latter two matrices can only describe the mapping relationship between nodal voltages and power injections. To solve an OPF that includes chance constraints on line flows, an extra but small effort is needed. Specifically, after the third step aforementioned, each ISO is able to express every nodal voltage (both magnitude and angle) as a function of the encrypted dispatchable generations. Since each line flow can be determined by the nodal voltages at both ends of the corresponding line, ISO $n$ can formulate the line flows within its area as functions of the encrypted dispatchable generations. After that, ISO $n$ can share these line flow expressions with others, which allows them to build the corresponding encrypted but equivalent line flow chance constraints. To protect line parameters from disclosure, ISO $n$ can randomly but equivalently modify the parameters in line flow expressions before sharing them.
% formulate the line flow constraint.

% 简要来说，因为线路潮流是由线路两端节点的电压幅值和电压相角决定的，而算法\ref{alg:PPD CCOPF}已实现节点电压幅值和电压相角的相关映射系数的计算，因此可以在这一基础上，进一步获得线路潮流的表达式，以及相应的约束。

% For details, please refer to Appendix. 

% \subsection{Characteristics of the Proposed Method}

% \subsubsection{Advantages}

The proposed distributed CC-OPF method has the following three main advantages:

1) The method is fully distributed. Only neighborhood communications between adjacent ISOs are required. This is because all the calculations are either implemented individually by each ISO or performed with the aid of distributed algorithms. Such a distributed structure suits the multi-party environment, e.g., a MIG with multiple regional ISOs. 

2) The method strictly protects the confidential information of different regions, due to the multi-layer encryption introduced by the TE technique and the PP-AAC algorithm. This feature makes the proposed method appealing to regional ISOs, because the risk of disclosing confidential information is avoided. This can promote active and effective cooperation between areas.

3) The method does not require any manual parameter tuning. Besides, unlike traditional distributed optimization methods, convergence is not a concern for the proposed method, as every ISO eventually solves a complete problem (encrypted though) instead of a sub-problem. It should be emphasized that, although the AAC and PP-AAC algorithms embedded in the proposed method require iterative calculations, they have robust convergence; most importantly, their parameters can be directly set via generalized rules.

\vspace{-0.2cm}
\section{Case Study}

This section verifies the performance of the proposed methods, including the fast distributed method for solving linear equation systems and the distributed CC-OPF method with confidentiality-preservation. Since their confidentiality-preserving features have already been discussed earlier and guaranteed theoretically, this section mainly verifies the accuracy and efficiency of these two methods.

% \subsection{Settings}
In the following, the IEEE 39-bus system ($N_a=5$) and the IEEE 118-bus system ($N_a=9$) are used for testing. In each test system, the communication topology among regional ISOs is a ring, i.e., each ISO only has two neighbors in terms of information exchange. Wind farms have been added to both test systems, and the joint probability distribution of wind power is assumed to be known.

Besides, the risk limit $\epsilon_{b}$ in \eqref{eq:supply-demand_dis} is set as $10^{-4}$. The chance constraints in \eqref{eq:state} are line flow chance constraints; the corresponding lines are those who directly link to wind farms. The probability limit of \eqref{eq:state} is set to $1-\alpha_S = 0.95$. The parameters of the AAC algorithm are the same as those in \cite{aysal2008accelerated}, while the extra parameters in the PP-AAC algorithm are specified as $\sigma = 2$ and $\gamma = 0.4$. 

All the experiments are implemented and run on a personal laptop with i5-7267u processor and 8 GB memory. The software environment is MATLAB 2017A, and the optimization solver is the  built-in function $\rm quadprog(\cdot) $ that calls the interior point method.

\subsection{Verifications of the Fast Distributed Method for Solving Linear Equation Systems}

For accuracy and efficiency verifications, the proposed fast distributed method is compared with a state-of-the-art approach, namely the accelerated projection-based consensus (APC) method \cite{azizan2019distributed}. The APC method has been proven to outperform a number of distributed approaches of the same type, including the distributed gradient descent, distributed versions of Nesterov’s accelerated gradient descent and heavy-ball method, the block Cimmino method, and ADMM \cite{azizan2019distributed}. 

For comparison, we do not use the two test systems introduced in the previous section but generate random square matrices of different dimensions (e.g., 45, 90, 135, 180), representing different sizes of the matrix G in order to simulate the performance for matrices of different scales. We assume nine parties, i.e. regional ISOs, and a ring as the communication topology. Each ISO only has information about partial rows of $\boldsymbol{G}$. For simplicity, the numbers of rows owned by different ISOs are assumed to be identical. These ISOs respectively use the proposed method and the APC method to compute the inverse of $\boldsymbol{G}$ and therefore obtain the desired sub-matrix from $\boldsymbol{G}^{-1}$, i.e., ISO $n$ obtains $\boldsymbol{B}_{Q,n}$ and $\boldsymbol{B}_{P,n}$. 
The average relative errors of these two methods compared to the true value of $\boldsymbol{G}^{-1}$ are given in Table \ref{tab:Accuracy_compare_APC}. Clearly, the proposed method outperforms the APC algorithm for these generic test cases. 
\vspace{-0.4cm}
\begin{table}[h]
	\caption{ Average Relative Errors and Computational Time (unit: seconds) of the Proposed Method and the APC Method}
	\label{tab:Accuracy_compare_APC}
	\centering
	\footnotesize
	\setlength{\tabcolsep}{1.9mm}{
	\begin{tabular}{c|cc|cc}
	\toprule
	\multirow{2}*{Dimension of $\boldsymbol{G}$}& \multicolumn{2}{c|}{\textbf{Error}} & \multicolumn{2}{c}{\textbf{Time }} \\
	~     & APC     & Proposed & APC     & Proposed       \\
	\midrule  
	45  & 4.78$\times 10^{-4}$ & 1.22$\times 10^{-11}$ & 5.75 & 0.08 \\
	90 & 8.74$\times 10^{-4}$ & 1.87$\times 10^{-11}$ & 25.17 & 0.14 \\
	135 & 3.34$\times 10^{-4}$ & 9.36$\times 10^{-12}$ & 182.48 & 0.24 \\
	180  & 7.18$\times 10^{-4}$ & 3.49$\times 10^{-9}$ & 295.79 & 0.41\\
	\bottomrule
	\end{tabular}}
\end{table}

% \begin{table}[h]
% 	\caption{ Computational Time of the Proposed Method and the APC Method (unit: second)}
% 	\label{tab:Efficiency_compare_APC}
% 	\centering
% 	\footnotesize
% 	\setlength{\tabcolsep}{1.2mm}{
% 	\begin{tabular}{cccc}
% 	\toprule
% 	Dimension of $\boldsymbol{G}$     & APC Method    & Proposed Method      \\
% 	\midrule
% 	45  & 5.75 & 0.08  \\
% 	90  & 25.17 & 0.14  \\
% 	135 & 182.48 & 0.24  \\
% 	180  & 295.79 & 0.41 \\
% 	\bottomrule
% 	\end{tabular}}
% \end{table}

It should be mentioned that in the above comparison, the stopping criterion for the APC method is chosen such that if the maximal deviation between the updated values in two adjacent iterations is lower than $10^{-4}$, then the iteration stops. But this threshold is set as $10^{-10}$ for the AAC and PP-AAC algorithms. We did not deliberately set a greater threshold for the APC method. In fact, if the stopping criterion is chosen smaller, the APC algorithm requires an unacceptable number of iterations, e.g. for $10^{-5}$ it will only stop after 20,000 iterations with long run-time. Also, even if we set the threshold for the AAC and PP-AAC algorithms to $10^{-4}$, the same as the APC, the relative error of the proposed method is $3.35\times 10^{-6}$ when the dimension of $\boldsymbol{G}$ is 180, which is still significantly less than the corresponding error of the APC algorithm. 

% In fact, we also want to set the threshold for the APC algorithm as $10^{-10}$

% We did not deliberately set the threshold of Method A to be larger

% The stopping criterion for the APC method is  chosen such that if the maximal deviation between the updated values in two adjacent iterations is lower than $10^{-4}$, then the iteration stops. The average relative errors of these two methods compared to the true value of G^-1 are given in Table II. If the stopping criterion is chosen smaller, the APC requires an unacceptable number of iterations, e.g. for 10^-5 it will only stop after 20'000 iterations. Hence, the proposed clearly outperforms the APC algorithm for these generic test cases.

% Choosing $10^{-4}$ is not intended to weaken the accuracy of the APC method, but to keep the computational burden  relatively reasonable, because if the tolerance is further reduced, say $10^{-5}$, then the APC method cannot stop within 20,000 iterations, which will lead to unbearable calculation cost.

Table \ref{tab:Accuracy_compare_APC} also provides the computational time of the evaluated methods. As can be seen, even if the stopping tolerance of the APC algorithm is set to $10^{-4}$, the time consumed by the APC method is still by dimensions longer than the time required by the proposed method, independent of the size of $\boldsymbol{G}$.

% To summarize, the proposed fast distributed method for solving linear equation systems has better accuracy and efficiency than the APC algorithm, thus outperforming a number of 
% state-of-the-art approaches that are not as good as the APC method.

\subsection{Verifications of the Distributed CC-OPF Method}

To verify the accuracy of the proposed distributed CC-OPF method, both  IEEE test cases described in Section VI-A are used, where $T$ for each case is set to 24 to simulate the day-ahead dispatch. We compare the results of the proposed method to its benchmark, i.e., solving $(\rm P0)$ in a centralized manner with access to the global information. 

\begin{figure}[h]
	\centering  
	\includegraphics[width=3in]{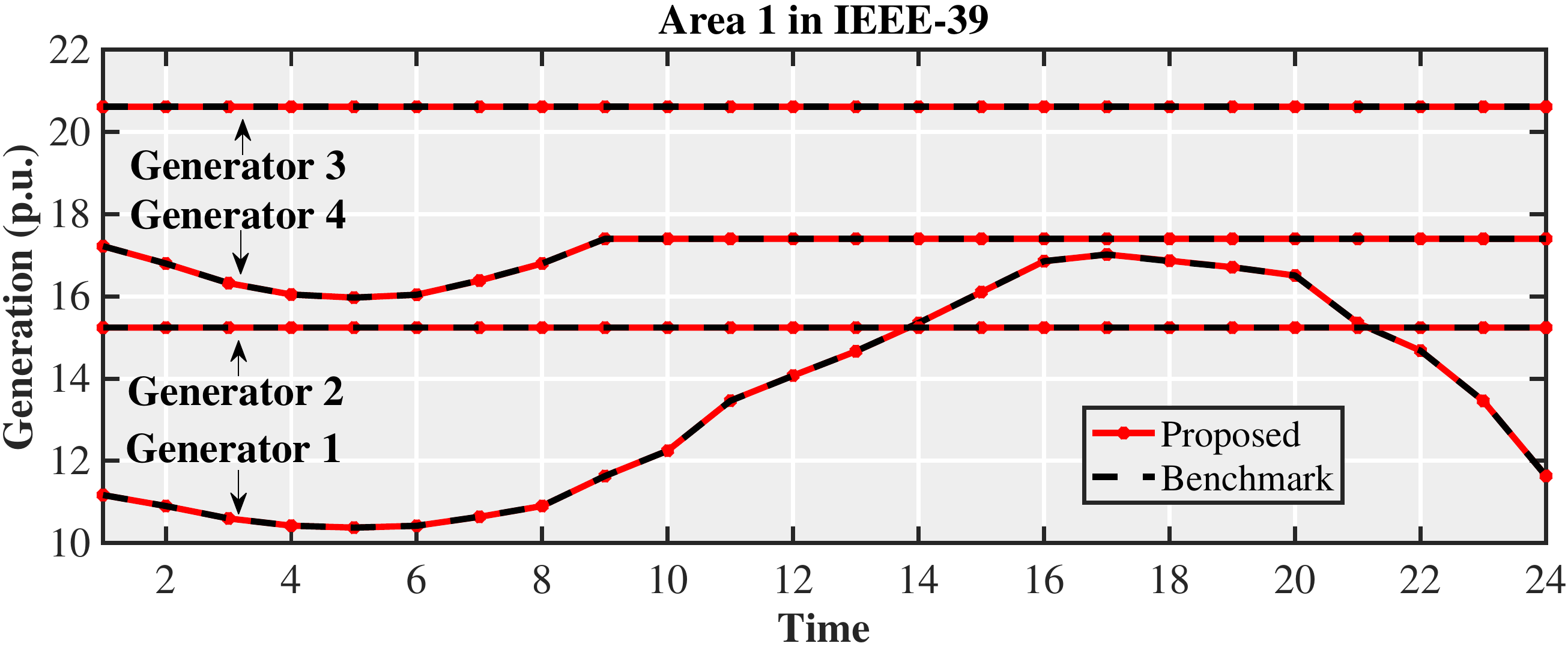} 
	\caption{Generator outputs of area 1 obtained by the proposed method and the benchmark method }
	\label{fig:generation} 
\end{figure}

Fig. \ref{fig:generation} shows the generator outputs obtained by the proposed and benchmark methods. Due to space limitation, only the generator outputs in area 1 from the IEEE 39-bus system is shown in this figure, where generators 2 and 3 reach their maximal output from the very beginning and generator 4 hits its maximum from $t=9$. As can be observed, the optimal solutions computed by the proposed method ideally coincide with the corresponding benchmarks. 
\vspace{-0.7cm}
\begin{table}[h]
	\caption{Objective Value Obtained by the Proposed Method and the Benchmark Method}
	\label{tab:obj_compare}
	\centering
	\footnotesize
	\setlength{\tabcolsep}{1.2mm}{
	\begin{tabular}{lccccc}
	\toprule
	Test Case     & Benchmark    & Proposed Method    & Relative Error    \\
	\midrule
IEEE-39  & 110958.06 & 110958.13 & 6.03$\times 10^{-7}$ \\
IEEE-118 & 8799.39124 & 8799.39122 & 2.12$\times 10^{-9}$  \\
	\bottomrule
	\end{tabular}}
\end{table}
\begin{table*}[h]
	\caption{Efficiency Comparison between the Proposed Method and Its Benchmark (unit: second)}
	\label{tab:time_compare}
	\centering
	\footnotesize
	\setlength{\tabcolsep}{1.5mm}{
	\begin{tabular}{ccccccccc}
	\toprule
	\multicolumn{1}{c}{\multirow{2}{*}{Test Case}} & \multicolumn{1}{c}{\multirow{2}{*}{$N_a$}} & \multicolumn{3}{c}{\textbf{Benchmark}} & \multicolumn{4}{c}{\textbf{Proposed Method}}     \\
	\multicolumn{1}{c}{}                      & \multicolumn{1}{c}{}                      &  Formulating      & Solving       & Total      & Formulating and Encrypting & Sharing     & Solving and Decrypting  & Total     \\
	\midrule
	IEEE-39                                   & 5                                         & 0.74    & 1.38     & 2.12    & 5.93  & 1.72   & 3.29   & 10.94  \\
	IEEE-118                                  & 9                                         & 6.11    & 24.84    & 30.95   & 51.16 & 45.78 & 145.07 & 242.01 \\
	\bottomrule
	\end{tabular}}
\end{table*}

Further, the optimal objective function values obtained by the proposed method and its benchmark are listed in Table \ref{tab:obj_compare} for both test cases. To avoid repetitive illustration, only the objective function values obtained by ISO 1 using the proposed method are shown. Table \ref{tab:obj_compare} indicates that, the proposed method has very high accuracy --- yielding relative errors of less than $7\times 10^{-7}$. These small deviations might be all numerical, yet they might also be caused by other possible reasons, e.g., 
\begin{enumerate}
	\item Although the encrypted problem $(\rm P1)$ is proved to be equivalent to $(\rm P0)$ mathematically, they might be  different from the view of commercial solvers. For example, most of the constraints in $(\rm P0)$ are uncoupled, while the encryption tightly couples the constraints in $(\rm P1)$. As a result, the commercial solver can easily identify and eliminate redundant constraints in $(\rm P0)$ before solving it, but fails to identify redundant constraint in $(\rm P1)$. This may lead to different start points when solving $(\rm P0)$ and $(\rm P1)$, thereby possibly yielding slightly different final solutions.
	\item The information exchange between ISOs is realized by the consensus-based algorithms, whose asymptotic convergence properties may introduce small error into the final results.
\end{enumerate}

% Note that the above experiment uses the calculation result of ISO 1 to represent the accuracy of the proposed method. In fact, the results obtained by other ISOs using the proposed method are also accurate. For illustration, Fig shows the optimal dispatchable generations computed by all ISOs, either using the proposed method or using the benchmark method. As

To verify the efficiency of the proposed distributed CC-OPF method, the computational time of each step in this method is measured and summarized in Table \ref{tab:time_compare}. Note that for the formulating, encrypting, and sharing steps, the time indicated is the overall computational time incurred by all ISOs, as they need cooperation to accomplish these steps. However, for the solving and decrypting steps, the time indicated is the maximal computational time incurred by an individual ISO, because these steps can be conducted by each ISO independently. The time consumed by the benchmark method is also listed in Table \ref{tab:time_compare}. This Table shows that the total computational time of the proposed method is always greater than that of its benchmark. Apparently, this is the price for enabling the distributed and confidentiality-preserving features. Nevertheless, the resulting overhead is acceptable for day-ahead dispatches.

% As the proposed method, solving and decrypting steps are the most time-consuming part, and far longer than the time spent in the solving step of the benchmark method. This is because the commercial solver eliminates a number of redundant constraints when solving $(\rm P0)$, but eliminates none constraint while solving $(\rm P1)$.

% Overall, 

% It is worth mentioning that, the larger the scale of the system, the more information each ISO needs to share. Hence in the large test case, e.g., the IEEE 118-bus system, the time spent by the sharing step is longer than that costed in the formulating and encrypting steps. 

% \subsection{Confidentiality-preservation Verification}
% The way used for preserving confidentiality in the two proposed methods are essentially the same: using random matrix to hide the real solution (either the solution of a linear-equations-system or a CC-OPF problem) via linear algebraic computation. Therefore, to avoid repetitive verification, only the confidentiality preserving feature of the proposed distributed CC-OPF method is tested here. 

% Concretely, 

\section{Conclusion}
This paper proposes a confidentiality-preserving distributed CC-OPF method on the basis of the TE technique. To this end, this paper first proves the TE technique's adaptability to quadratic programming. Then, a fast distributed method for solving linear equation systems is developed, which enables regional ISOs to compute the coupled coefficients in chance constraints when they do not have the global information. Finally, the distributed CC-OPF method with confidentiality preservation is proposed. This distributed method can ensure that regional ISOs obtain and only obtain the accurate optimal dispatchable generations within their own regions without disclosing their confidential data.

\bibliographystyle{IEEEtran}
\bibliography{IEEEabrv,paper}
\end{document}